\documentclass[sn-mathphys,iicol,Numbered]{sn-jnl}

\usepackage{graphicx}%
\usepackage{multirow}%
\usepackage{amsmath,amssymb,amsfonts}%
\usepackage{amsthm}%
\usepackage{mathrsfs}%
\usepackage[title]{appendix}%
\usepackage{xcolor}%
\usepackage{textcomp}%
\usepackage{manyfoot}%
\usepackage{booktabs}%
\usepackage{algorithm}%
\usepackage{algorithmicx}%
\usepackage{algpseudocode}%
\usepackage{listings}%
\usepackage{siunitx}

\theoremstyle{thmstyleone}%
\theoremstyle{thmstyletwo}%

\theoremstyle{thmstylethree}%

\begin{document}

\title[Article Title]{Demonstration of magnetically silent optically pumped magnetometers for the TUCAN electric dipole moment experiment}


\author*[1]{\fnm{Wolfgang} \sur{Klassen}}\email{wklassen@triumf.ca}

\author[3,4]{\fnm{Shomi} \sur{Ahmed}}

\author[3,6]{\fnm{Kiera} \sur{Pond Grehan}}

\author[5]{\fnm{Chris} \sur{Hovde}}

\author[1]{\fnm{Kirk W.} \sur{Madison}}

\author[3,4]{\fnm{Russell R.} \sur{Mammei}}

\author[3,4]{\fnm{Jeffery W.} \sur{Martin}}

\author[3]{\fnm{Mark} \sur{McCrea}}

\author[4]{\fnm{Tahereh}  \sur{Mohammadi}}

\author[2]{\fnm{Takamasa} \sur{Momose}}

\author[5]{\fnm{Patrick} \sur{Opsahl}}

\author[3]{\fnm{David C.M.} \sur{Ostapchuk}}

\affil*[1]{\orgdiv{Physics and Astronomy}, \orgname{University of British Columbia}, \orgaddress{\street{6224 Agricultural Rd}, \city{Vancouver}, \postcode{V6T 1Z1}, \state{British Columbia}, \country{Canada}}}

\affil[2]{\orgdiv{Chemistry}, \orgname{University of British Columbia}, \orgaddress{\street{2036 Main Mall}, \city{Vancouver}, \postcode{V6T 1Z1}, \state{British Columbia}, \country{Canada}}}

\affil[3]{\orgdiv{Physics}, \orgname{University of Winnipeg}, \orgaddress{\street{515 Portage Avenue}, \city{Winnipeg}, \postcode{R3B 2E9}, \state{Manitoba}, \country{Canada}}}

\affil[4]{\orgdiv{Physics and Astronomy}, \orgname{University of Manitoba}, \orgaddress{\street{30A Sifton Road}, \city{Winnipeg}, \postcode{R3T 2N2}, \state{Manitoba}, \country{Canada}}}

\affil[5]{\orgname{Southwest Sciences}, \orgaddress{\street{1570 Pacheco Street}, \city{Santa Fe}, \postcode{87505}, \state{New Mexico}, \country{United States}}}

\affil[6]{\orgdiv{Department of Physics and CQIQC}, \orgname{University of Toronto}, \orgaddress{\city{Toronto}, \postcode{M5S 1A7}, \state{Ontario}, \country{Canada}}}


\abstract{We report the performance of a magnetically silent optically pumped cesium magnetometer with a statistical sensitivity of 3.5~pT/$\sqrt{\mathrm{Hz}}$ at 1~Hz and a stability of 90 fT over 150 seconds of measurement. Optical pumping with coherent, linearly-polarized, resonant light leads to a relatively long-lived polarized ground state of the cesium vapour contained in a measurement cell.  The state precesses at its Larmor frequency in the magnetic field to be measured.  Nonlinear magneto-optical rotation then leads to the rotation of the plane of polarization of a linearly polarized probe laser beam.  The rotation angle is modulated at twice the Larmor frequency.  A measurement of this frequency constitutes an absolute measurement of the magnetic field magnitude.  Featuring purely optical operation, non-magnetic construction, low noise floor, and high stability, this sensor will be used for the upcoming TUCAN electric dipole moment experiment and other highly sensitive magnetic applications.  Novel aspects of the system include commercial construction and the ability to operate up to 24 sensors on a single probe laser diode.}

\keywords{magnetometer, magnetometry, neutron electric dipole moment, all-optical, alkali, atomic vapour, magnetically silent, nonlinear magneto-optical rotation, Faraday rotation, Bell-Bloom, Cesium, free precession decay}

\maketitle

\section{Introduction}\label{sec1}
Experiments at the frontiers of precision low-energy measurements push the limits of existing measurement technology.  The search for a non-zero neutron electric dipole moment (nEDM) is pursued by several labs around the world~\cite{panEDM2019,n2EDM2019,lanl2018,pnpi_ILL2015,sns_2019,JPARC_2018}.  A non-zero nEDM would have a very high impact on physics beyond the standard model, and could help explain the strong charge-parity (CP) problem~\cite{carena_2019,Mimura_2019}, the baryon asymmetry observed in the universe~\cite{Morrissey_2012,nfbell_2019}, or sources of CP violation beyond the standard model~\cite{cirigliano_2019,Crivellin_Saturnino_2019}.  In particular, the sensitivity of the nEDM to strong sector physics makes this measurement attractive above other fundamental EDM searches such as the measurement of the electron EDM, which has been done to very high precision~\cite{acme_2018}.

The current best measurement of the nEDM is $d_n = 0.0\pm1.1_{\mathrm{stat}}\pm0.2_{\mathrm{sys}}\times10^{-26}$~$e\cdot\mathrm{cm}$~\cite{psi2020}, performed at the Paul Scherrer Institute (PSI) by the PSI nEDM collaboration.  The TRIUMF Ultracold Advanced Neutron (TUCAN) collaboration~\cite{tucanTRIUMF} aims to achieve an experimental sensitivity of $10^{-27}$ $e\cdot\mathrm{cm}$ using a new spallation-driven superfluid helium ultracold neutron (UCN) source at TRIUMF~\cite{TUCAN_firstucn,TUCAN_kicker,TUCAN_beam,TUCAN_moderator}. 

Measurements of the nEDM require extremely well measured and controlled magnetic fields~\cite{abel2020,psimapping_2022,psi2020}.  Cs magnetometry was used in conjunction with Hg comagnetometry in the latest upper limit paper~\cite{psi2020}.  The Cs sensors used by the PSI nEDM collaboration were demonstrated to have sensitivities of \mbox{0.75-8}~pT/$\sqrt{\mathrm{Hz}}$ and absolute accuracies of \mbox{45-90}~pT per measurement~\cite{abel2020}.  However, due to the method by which these sensors are operated they are not magnetically silent (Sec.~\ref{sec:config}), and can therefore not be used while neutrons are being measured and thus have limited utility.

Over the past several decades an alternate method~\cite{Budker2002} of creating and probing polarized alkali atoms has been developed that uses non-linear magneto-optical rotation (NMOR), which allows for the development of robust, drift-stable, completely magnetically silent optical magnetometers~\cite{Gawlik2008,Das_2018}.

In this work we demonstrate a statistical sensitivity of 3.5~pT/$\sqrt{\mathrm{Hz}}$ at 1~Hz and a stability of 90 fT over 150 seconds of measurement in a configuration that operates purely optically, using NMOR to both create and probe a polarized atomic state.  This allows precision magnetic field measurements to be done during the EDM experiment without perturbing the neutrons~\cite{klassen2020}.
Novel aspects of the design include completely commercial components (Sec. \ref{sec:config}) and the ability to run many sensors on a single laser diode (Sec. \ref{sec:results}).

\section{Configuration}\label{sec:config}

\begin{figure}
    \centering
    \includegraphics[width=0.5\textwidth]{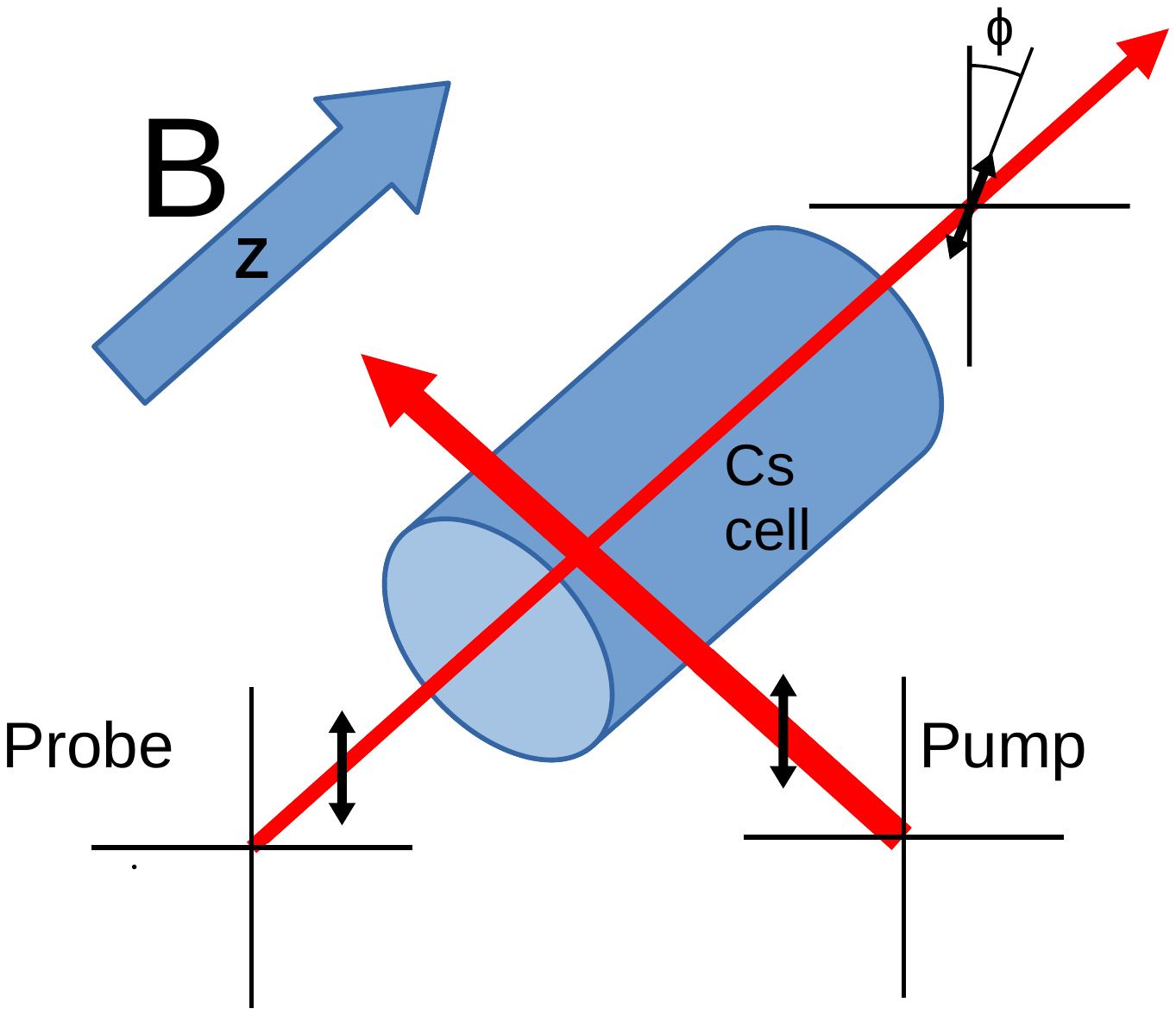}
    \caption{Shown is a schematic of the vapour cell with the orientation of the magnetic field and optical beams used for this magnetometer.  Both the pump and probe beams are linearly polarized with their electric field vectors aligned and perpendicular to the magnetic field direction.  The probe beam propagates through the vapour cell along the magnetic field vector to be measured, while the pump beam can be anywhere in the plane formed by the two beams in this figure.  The magnetic field information is encoded in the time dependence of the polarization angle $\phi$ of the probe beam after it exits the vapour cell.}
    \label{fig:pumpscheme}
\end{figure}

Atomic magnetometry involves interrogating the Larmor frequency of atoms exposed to a magnetic field.  Generally, this involves the creation of a large coherently precessing population of atoms in some relatively long-lived polarization state.  The polarized atoms then precess in the magnetic field they experience.  The precession is probed optically using resonant light, and the measured magnetically sensitive quantity is the Larmor frequency of the polarized atoms~\cite{Budker2002,higbie2006}.

We have chosen a mode of operation which operates similarly to a Bell-Bloom configuration~\cite{bellbloom}, with some modifications to bring the operation away from zero field.  This follows work done by Higbie et. al.~\cite{higbie2006} developing NMOR based alkali magnetometry and also work done by Gru\'jic et. al.~\cite{Grujic_2015} in developing the FSP mode of operation.

In order to illustrate the advantages of this configuration we first present a brief overview of the sensors used in the PSI nEDM collaboration's latest upper limit paper~\cite{psi2020}, which operate in an $M_x$ configuration~\cite{abel2020}, similar to some commercially available alkali magnetometers.  Circularly polarized light (CPL) propagates at a 45$^\circ$ angle to the measured field, creating the polarized population along the direction of propagation.  A radio frequency (RF) magnetic field resonant with the polarized state drives the polarization into the plane transverse to the magnetic field, causing the state to precess.  The absorption of the CPL is proportional to the transverse magnetization of the atoms, so the Larmor frequency is accessible via monitoring the intensity of the CPL after it passes through the cell.  The same beam is used for both tasks.  In the main configuration used during nEDM measurements the absorption signal is phase-shifted and fed back to the RF coils such that the system self-oscillates at the Larmor frequency of the atoms.  This phase is then proportional to the difference between the driving frequency and the larmor frequency of the atoms.  By locking this phase to 0, the frequency of the RF coil is identified as the Larmor frequency, and the field is measured.

This configuration is robust and high-bandwidth, but due to the RF it is not magnetically silent and is subject to two main types of error: errors in determining the Larmor frequency (i.e. errors due to drifting phase or electronic changes due to temperature changes) and systematic effects that change the Larmor frequency itself, making it an inaccurate measure of the relevant field.  It is also subject to potential long-range crosstalk due to the RF used in the sensors.

The PSI experiment avoided the first kind of error by using a Free Spin Precession (FSP) mode of operating.  Instead of feeding the absorption signal back to the RF coils, the atoms were allowed to freely precess after the RF pulse is applied, and the frequency of the resulting oscillating absorption signal was measured.  This is a phase-error-free measurement of the Larmor frequency, and it was used to verify the offsets present in the main mode of operation.  However, the FSP operation mode was not an option during nEDM measurement runs because the RF pulses used in this mode were repeated at close to the Larmor frequency of the mercury comagnetometer and would potentially cause interference.

PSI carefully studied and controlled the various shifts associated with the second kind of error: shifts of the Larmor frequency itself.  The largest effect was due to the AC Stark shift, which is caused by the component of CPL light propagating along the measured field.  This was found to cause shifts of $\pm$10~pT to $\pm$50~pT which were correlated with the light intensity and frequency~\cite{psi2020}.  These shifts were only observable using auxiliary offline measurements, so were not able to be constrained any further than this.

The Bell-Bloom style Free Precession Decay (FPD) configuration used in our system also measures the Larmor frequency of Cs to deduce the magnetic field experienced by the atoms; however, it uses nonlinear magneto-optical rotation~\cite{higbie2006,budker2013} rather than absorption to both polarize the atoms and interrogate their Larmor frequency. In this configuration, the electric field vector of both the pump and probe beams are perpendicular to the magnetic field, as in figure \ref{fig:pumpscheme}.  The linearly polarized light can be described as an equal mixture of left- and right-handed CPL, and so will induce $\mathrm{m}_{\mathrm{F'}} = \mathrm{m}_{\mathrm{F}}\pm1$ transitions.  The natural decay of the excited state has a random angular momentum transfer, and so on average the atoms are pushed towards states with $\mathrm{m}_{\mathrm{F}}=\pm$F.  Classically this corresponds to the spins of the atoms in the ensemble having a preferred axis of alignment, so this is called an ``aligned'' state, and the atoms can be said to have been polarized.  The magnetic field to be measured now exerts a net torque on these aligned spins, and the atoms precess in the magnetic field at their Larmor frequency.  The pump beam is quickly turned off to allow precession to occur.  After half of one Larmor period the alignment axis of the atoms is along the laser polarization axis once again, and the pump beam is switched back on, reinforcing the aligned state.
By repeating this process, a large population of such polarized atoms can be produced.  Once the population is maximized, the pump beam is switched off.  A separate, much weaker probe beam is then passed through the atoms, propagating parallel to the magnetic field.  The polarized state of the Cs exhibits an axis of birefringence that precesses along with the atoms.  This causes the angle of the linearly polarized probe light, $\phi$, to be modulated at twice the Larmor frequency of the atoms because of the twofold symmetry of the aligned state.  This modulation is detected via polarimetry after the probe beam exits the vapour.

The FPD configuration eliminates the issue of RF crosstalk and interference by eliminating the use of RF entirely.  The issue of phase drift is eliminated by running exclusively in FPD mode, while the issue of the vector light shift is reduced by using linearly polarized light for the probe beam.  Additionally, servicing all sensors with a single probe laser means any systematic shifts associated with laser diode intensity or frequency fluctuations should not impact relative measurements between sensors, meaning gradient extraction is not affected.  These same changes are planned by the PSI collaboration for their upcoming n2EDM experiment~\cite{n2EDM2019} and have been implemented by the panEDM collaboration as well~\cite{panEDM2019,rosner2022,Rosner2022thesis,Sturm:2020gip}.
The sensors used in our work were not made in-house, but rather by an externally contracted commercial manufacturer who assembled the sensors using entirely commercially available components.  This project would not have been possible without the involvement of the commercial vendor, Southwest Sciences~\cite{SouthwestSciences_2016}.  Using commercial components allows us to easily expand the system to accommodate more sensors (Sec. \ref{sec:results}).  The use of commercially purchased components and the decision to permanently epoxy the vapour cells in the sensor heads are the main features that distinguish our system from the panEDM system.
Using the BMSR2 at PTB in Berlin the panEDM team was able to produce an extremely stable field using SQUID feedback, and they demonstrated an impressive 600-700~fT average difference between successive measurements~\cite{rosner2022}.  This can most closely be compared to our Allan deviation at minimum integration time, which was 3.9 pT (Sec. \ref{sec:results}).  They also demonstrated a stability below 50~fT between 70 and 600~s.

\section{Apparatus}\label{sec:apparatus}

\begin{figure}
    \centering
    \includegraphics[width=0.5\textwidth]{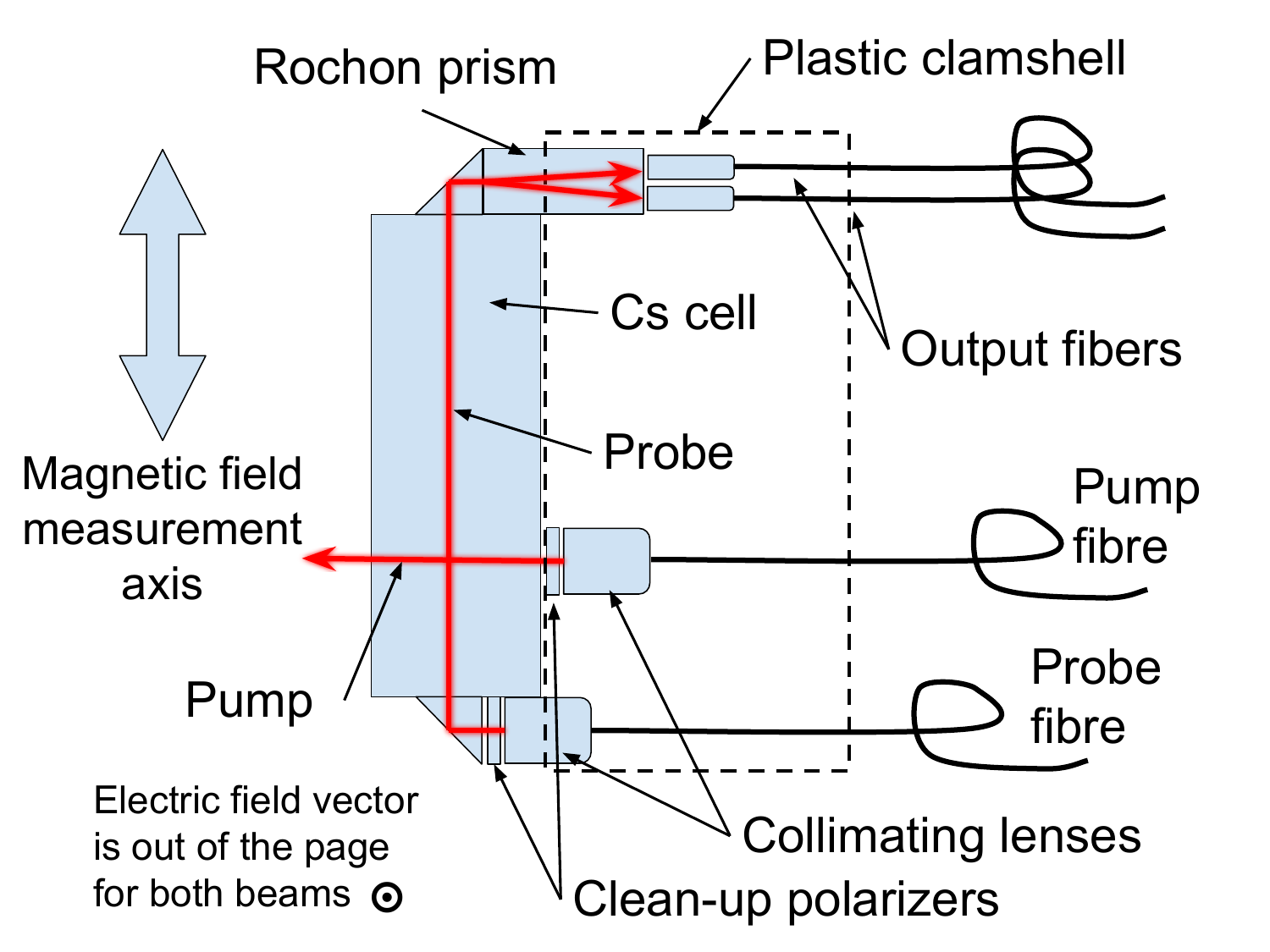}
    \caption{Schematic view of the sensor head, showing the optical components.  The two beams propagate orthogonally to each other, but their electric field vectors are aligned.  The most sensitive measurement axis for the magnetic field is indicated}
    \label{fig:sensorhead}
\end{figure}

\begin{figure}
    \centering
    \includegraphics[width=0.5\textwidth]{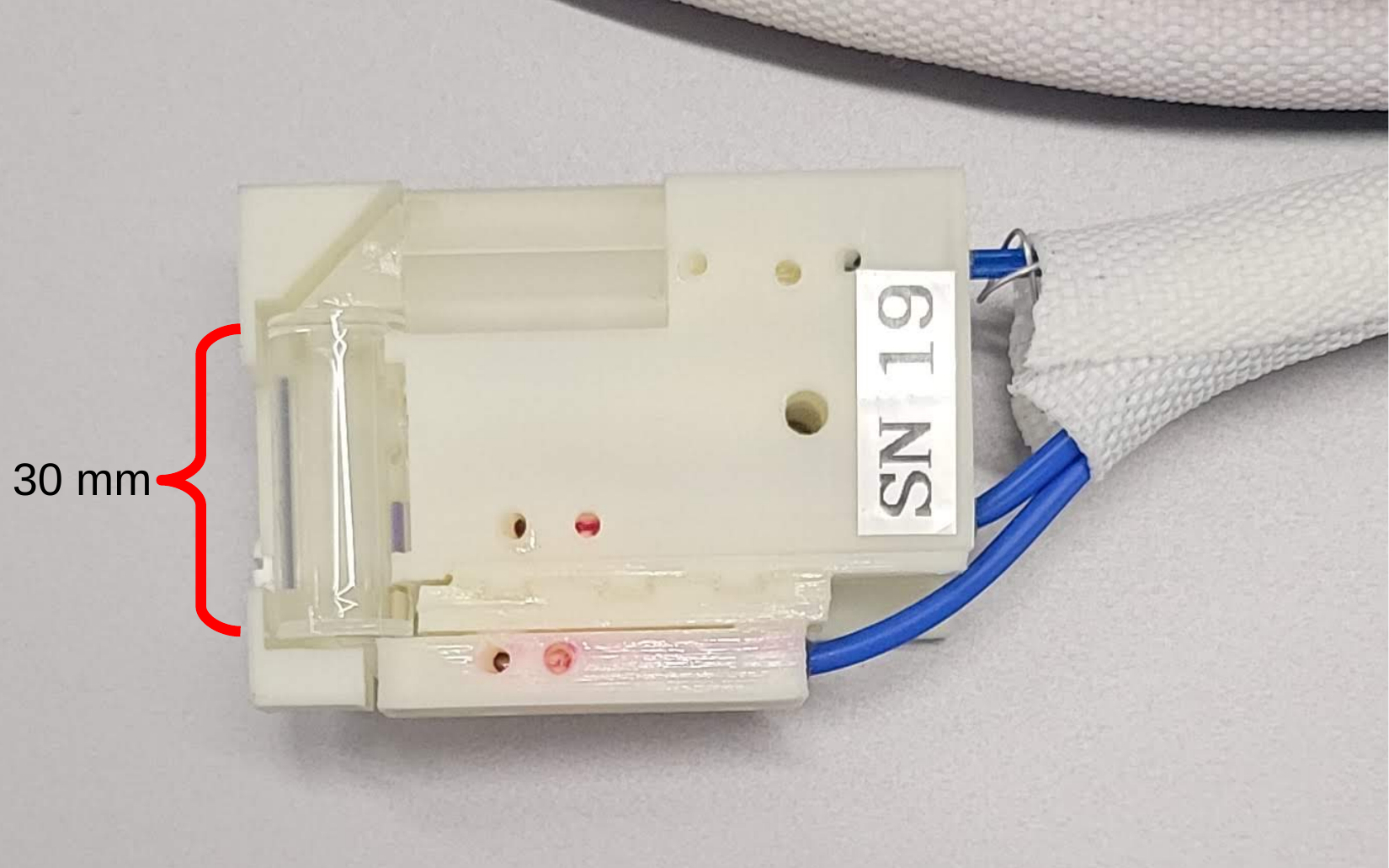}
    \caption{Cs sensor serial number 19, shown in the same orientation as the schematic diagram in figure \ref{fig:sensorhead}}
    \label{fig:sn19}
\end{figure}

\begin{figure}
    \centering
    \includegraphics[width=0.5\textwidth]{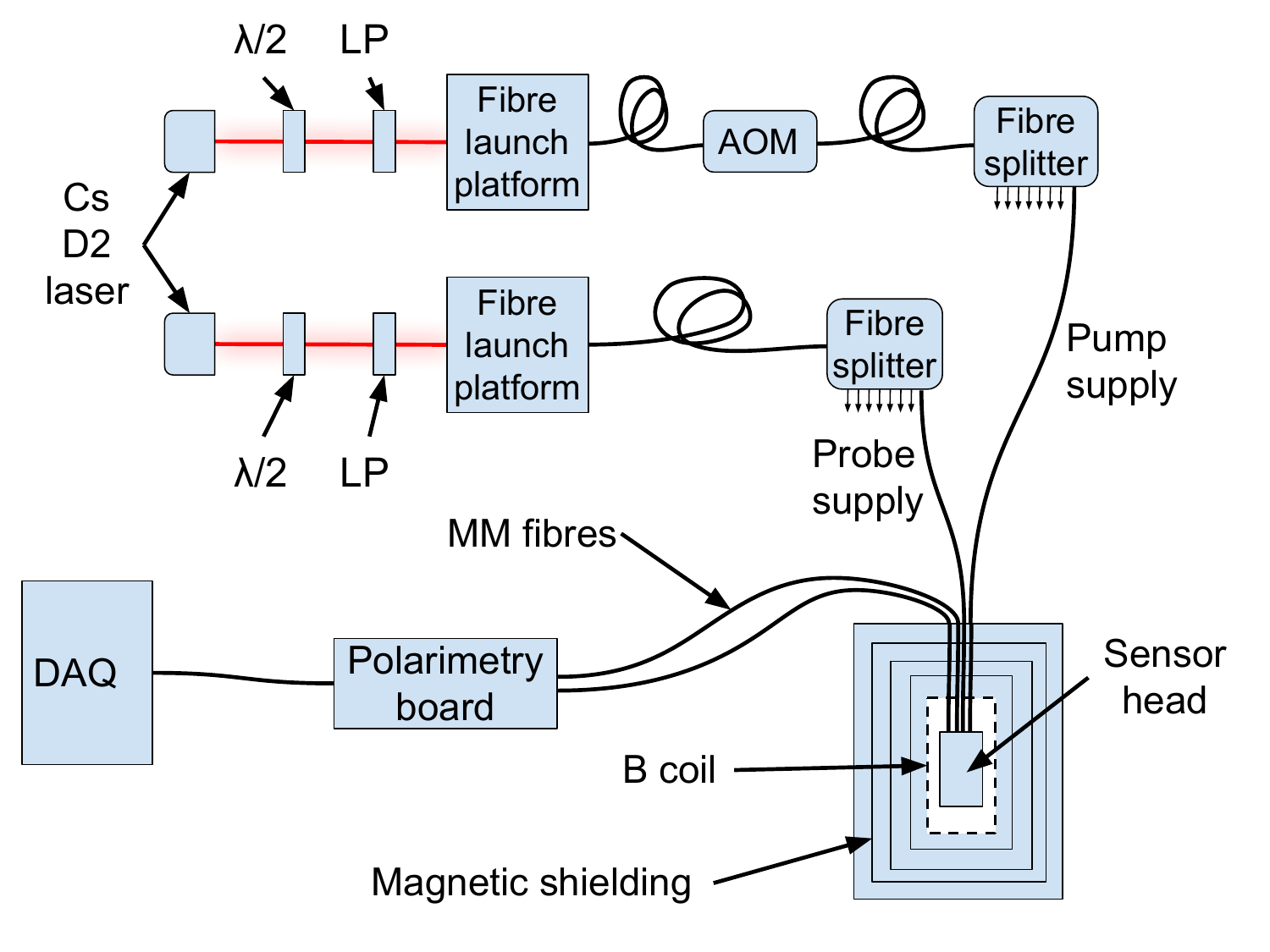}
    \caption{Schematic view of the optics that are not contained in the sensor head.  The free space optics, before the fibres, are bolted to an optical breadboard.  This configuration can operate 8 sensors simultaneously, although this has not been done due to the space limitations of our magnetically shielded volumes.  Labelled elements are a half-wave plate ($\lambda/2$), linear polarizer (LP), acousto-optic modulator (AOM),  multi-mode optical fibre (MM), and data-acquisition system (DAQ)}
    \label{fig:schematic}
\end{figure}

Figure \ref{fig:sensorhead} shows a schematic view of the optical components of the sensor head, figure \ref{fig:sn19} shows a picture of the sensor head in the same orientation, and figure \ref{fig:schematic} shows the physical setup and highlights the parts of the beam that are free space vs contained in fibre optics.  The laser diodes (indicated as Cs D2 laser in figure \ref{fig:schematic}) are held in precision motion stages on an optical breadboard.  Half wave plates and linear polarizers (indicated as $\lambda/2$ and LP respectively) are inserted into this free-space section of beam to control the polarization with which the beams are launched into the fibres.  After being launched into fibre optics the light is split so that multiple sensors can run on the same diode.  Fibre optics transport the light into the magnetically shielded environment.  After leaving the sensor head the light is carried back out of the magnetically shielded environment via large diameter multimode (MM) fibres.  The light from the MM fibres is detected via photodiodes mounted to the polarimetry boards, the signals of which are digitized by our DAQ system for analysis.

The laser diodes are EYP-DFB-0852-00150-1500-TOC03-0005 from Toptica Eagleyard.  They can be locked using a dichroic atomic vapour laser locking (DAVLL) system, although the diodes are stable enough to not require locking for short-term operation, on the order of hours.  The diodes are powered by highly stable diode current supplies from Vescent (D2-005 supply, D2-105-200 controller), and feedback control for the DAVLL is possible with Vescent laser lock boxes (D2-125 servo).  Locking was not needed for these studies, we verified the laser frequency carefully before and after each measurement to ensure it was stable by sweeping over the D2 absorption lines and locating our frequency relative to the absorption peaks.  The diodes are collimated into a $\sim$3~mm$^2$ free-space beam for a short length, into which polarization optics can be placed to specify launch polarization/intensity.  After this short free-space beam, the lasers are launched into fibre.

All fibre in this system before the sensor is either polarization maintaining (PM) or polarizing (PZ) fibre.  This ensures very clean linear polarization in the sensor head.  The pump laser is launched into the input fibre of a fibre-coupled acousto-optic modulator (AOM).  This is then attached to a $1\times8$ fibre splitter (also using PM fibre) which splits the pump light equally into 8 pigtailed PM fibres, each of which can provide pump light to a sensor.  The probe light is launched directly into another $1\times8$ PM splitter.  The overall intensity of the probe beams are controlled by a linear polarizer and half-wave plate in the free space section of the beam.  Individual ports can be additionally attenuated using an adjustment screw on the body of the splitter.  The free space beams are $\sim$100~mW, typical launch efficiencies with the current opto-mechanical setup are \mbox{20-30\%}, leading to \mbox{$\sim$1-2}~mW available for use in the sensor heads after accounting for all losses from splitting and attenuation.  This is sufficient power to optimize the pump light~\cite{klassen2020} and more than enough to provide probe power.  The pump beams can be globally attenuated by adjusting the high level of the square wave operating the AOM.  The probe beams are typically $<5~\mu$W in the sensor head after being adjusted using screw-attenuators built into the fibre splitter output ports.  $<5~\mu$W is enough to reach shot-noise-limited performance in the photodiodes without limiting the coherence time of the Cs via de-pumping.  The exact power is optimized to minimize the uncertainty in the frequency fit of the FPD signal.

The magnetically sensitive component of these sensors is an evacuated glass cell with a small amount of Cs held in a reservoir, attached to the cell via a capillary tube.  This cell was manufactured by Precision Glass Blowing (PGB)~\cite{PGB2024}.
The cell is a cylinder 30~mm in length and 10~mm in diameter.
The cell is treated with an anti-relaxation alkene coating available from PGB prior to being filled, which greatly extends the polarization lifetime of the atoms in the cell by reducing polarization-destroying wall collisions.
Prior to being sent to be installed by the contracted manufacturer the important atomic properties of the cell --- the coherence time, vapour density etc. --- were evaluated at the University of Winnipeg optics lab.  It was possible via careful heat treatment and handling to improve these properties.  Ultimately just over half of the cells purchased from PGB were determined to be of high enough quality to be installed in sensors.
Light is delivered to the Cs cell where it is collimated into a free-space beam to pass through the cell and interact with the Cs.  After passing though the Cs cell the probe beam is directed through a polarizing beamsplitter and the resulting two beams are launched into MM fibre to be carried out of the magnetically sensitive environment for detection and analysis.
This sensor head was designed and manufactured by Southwest Sciences~\cite{SouthwestSciences_2016} in consultation with TUCAN, and incorporates design improvements developed in Munich~\cite{panEDM2019,rosner2022,Rosner2022thesis,Sturm:2020gip}, mainly the use of the clamshell design, which will now be described.

The optical components are held in a plastic clamshell with a very similar design to the current panEDM Cs design~\cite{rosner2022}, 3D printed in undyed ABS plastic.
The long axis of the cell is transverse to the direction of the incoming fibre.  This results in the sensitive axis of the sensor being transverse to the long dimension of the sensor head (Fig.~\ref{fig:sn19}).
Optical components are cemented to one another at the optical interfaces using RD3-74 epoxy~\cite{Summersoptical}.  The epoxy provides the structural connection and maintains alignment between optical components, with the clamshell providing a mounting point for the whole sensor.
In the first TUCAN experimental phase these sensors will not be operated in vacuum, so the vacuum compatibility of these materials has not been assessed, however we expect that the components used are vacuum compatible.
The optical epoxy ensures efficient light extraction from the sensor head.
Linearly polarized light from two laser diodes is delivered to the sensor head via polarizing fibre optics.  A gradient-index (GRIN) lens launches light from the probe fibre to a $\sim$1~mm$^2$ free-space beam, which is then directed through a clean-up linear polarizer, followed by the vapour cell via internal reflection through a right-angled prism.  This probe launching assembly is mounted to the main ABS body using 
a coupling which constrains the motion of the probe assembly to linear motion in a plane, which helps with alignment.
Linear translation stages are then used to position the assembly to maximize coupling into the exit fibers prior to cementing the prism to the face of the vapour cell as the final step in optical alignment.  The beam is then again redirected via a right-angled prism through a Rochon walk-off polarizing beam splitter which splits the beam into orthogonal, linearly polarized components.  These components are then launched back into fibre optics: much larger core MM fibres are used on this side so as to collect the most light possible.  The two polarization components are then used for polarimetry of the probe light.  A GRIN lens also launches linearly polarized pump light into the cell after passing it through a clean-up polarizer, due to the geometry of our setup it can pass straight through the cell at 90$^\circ$ to the probe beam, since the electric field vectors of the two beams are aligned.  This choice of beam geometry was largely driven by considering ease of manufacture, launching both beams into the end of the vapour cell would be difficult to do.  Any loss in pumping efficiency due to the shorter optical path length can be compensated with pump power.

\begin{figure}
    \centering
    \includegraphics[width=0.5\textwidth]{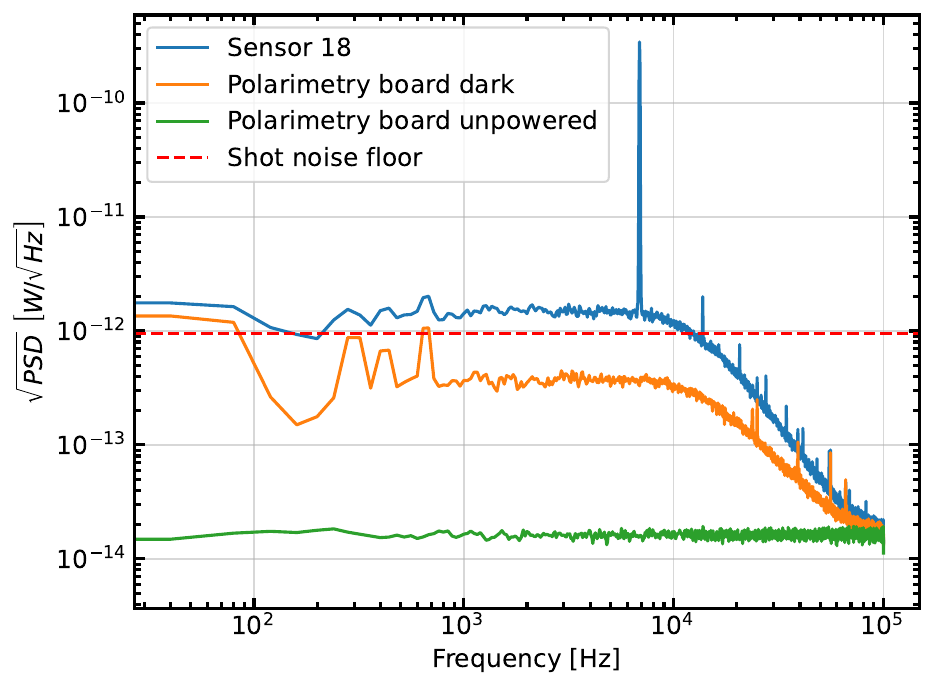}
    \caption{Square root of the power spectral density (PSD) of a full second of the unfiltered $\sim$~7kHz sinusoid measured by the polarimetry board.  The 7~kHz is generated by continuously flashing the pump beam at the Larmor frequency of the cesium, and the sensor is held at 1~$\mu$T for the test.  The total power of the probe beam was 3.9~$\mu$W, shot noise limit shown as the dotted line.  The average noise power in the frequency band of interest is 1.4~pW/$\sqrt{\mathrm{Hz}}$, and the amplitude of the sinusoid is 3~nW for a signal to noise ratio (SNR) of 2142}
    \label{fig:PSDshotnoise}
\end{figure}

The Rochon prism is oriented with its fast-axis at a 45$^\circ$ angle relative to the probe beam polarization axis so that it evenly splits the unperturbed probe light into two components of equal power.  This means that for small rotations, the degree of rotation of polarized light is directly proportional to the difference in power between the two components.  The MM fibres are plugged into a polarimetry board, which amplifies the photo-current difference from the two photodiodes watching the two fibres.  The resulting signal is then captured on an oscilloscope, or via a 32 channel DAQ system made by D-tAqc~\cite{D-TACQ}.  The D-tAqc system has a bandwidth of 200~kHz with 16 bit resolution.  The polarimetry board differencing circuit has a high-frequency cutoff of around 15~kHz, and is AC coupled since we are not interested in static imbalances in the two polarization channels.  This relatively low frequency cut-off was due to regulatory requirements imposed on the manufacturer rather than any physics motivation.
Figure \ref{fig:PSDshotnoise} compares the signal from the board in three states; unpowered, dark, and with the pump beam continuously driving oscillation at the Larmor frequency.  Each of these signals is integrated for a full second.  The dark noise of the polarimetry board is below the shot-noise limit for the power of the probe beam, meaning the electronics are capable of shot-noise limited detection of the signal.

\begin{figure}
    \centering
    \includegraphics[width = 0.5\textwidth]{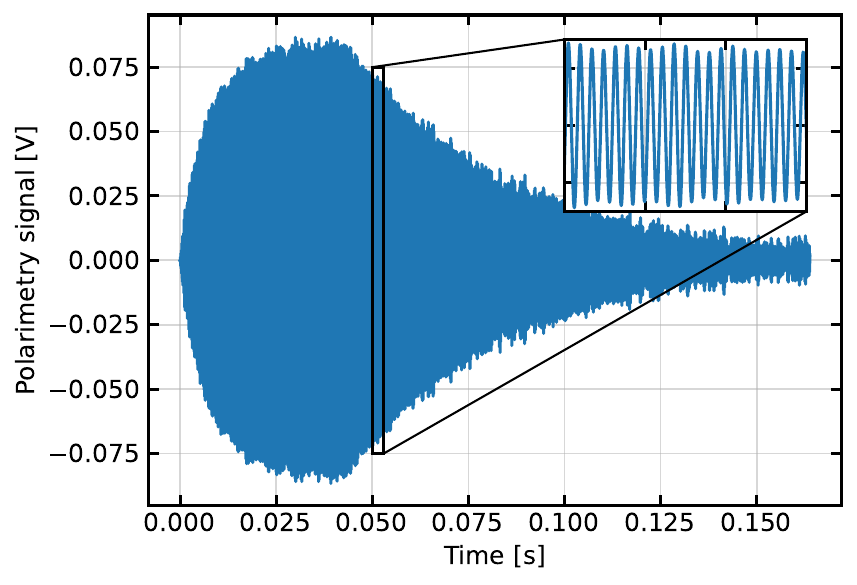}
    \caption{The amplified, differential photo-current corresponding to the degree of rotation of polarized light as a function of time during the polarization and the free spin precession phases of the Cs gas.  Pumping begins at T=0 on this plot, and stops just after 40~ms when the signal reaches its maximum.  During this time the pump beam is modulated into a 20\% duty cycle square wave at $\sim$7~kHz, approximately the Larmor frequency of Cs in a 1~$\mu$T field.  The remainder of the signal is the free spin precession portion which is later fit to an exponentially decaying sine wave.  A single instance of this pump-probe cycle is sometimes referred to as one free precession decay cycle, or one ``FPD'' cycle}
    \label{fig:singleFID}
\end{figure}

\begin{figure}[t]
    \centering
    \includegraphics[width=0.5\textwidth]{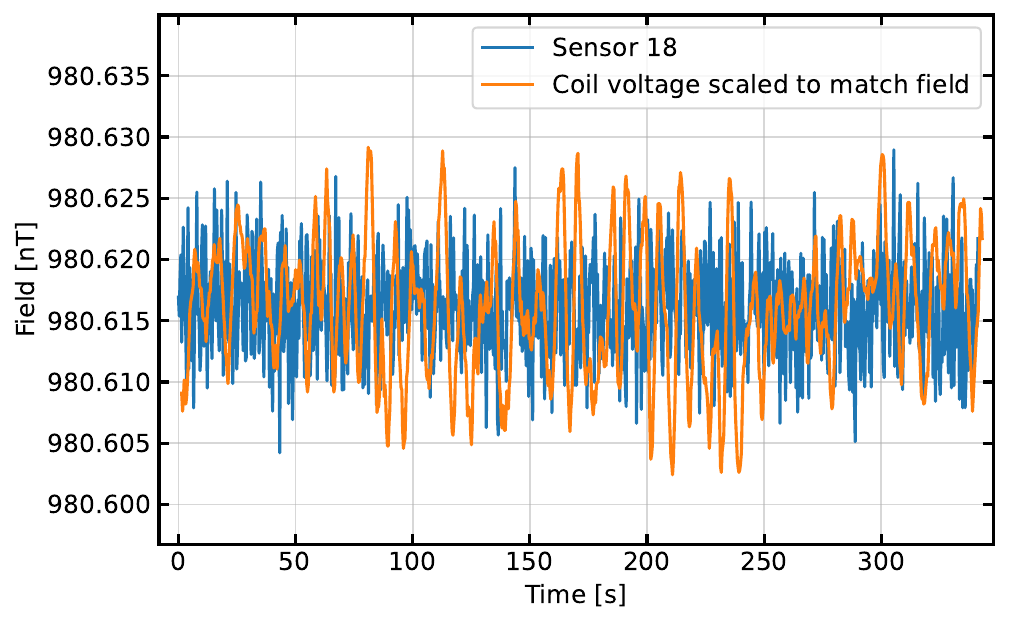}
    \caption{The measured magnetic field (blue line) and measured coil voltage (orange line) versus time.  The voltage across is being monitored by an 8.5 digit multimeter from Keithley, which is set to an integration time of 4/60 of a  second, or 4 power line cycles, matching the integration time of the FPD signal measuring the magnetic field
    \label{fig:samplemeasurement}}
\end{figure}

 The probe laser frequency is tuned near the Doppler broadened \mbox{F=4 $\rightarrow$ F$'$=3,4,5} D2 absorption line of Cs at 852~nm.  The pump laser is tuned to the Doppler broadened \mbox{F=3 $\rightarrow$ F$'$=2,3,4} absorption line.  The particular frequencies are chosen to maximize the magnetometry signal.  The pump is adjusted until a maximally polarized ensemble is reached.  The probe adjustment is a trade-off between amplitude and $T_2$.  Generally a high amplitude probe signal is correlated with a poor $T_2$, and vice versa.  The frequency is adjusted until the area under the decay envelope is maximized, which results in the best fit uncertainty.  This envelope is visible in figure \ref{fig:singleFID}.  Typically the best probe laser frequency is found around 200~MHz below the \mbox{F=4 $\rightarrow$ F$'$=3,4,5} absorption peak.  The best pump laser frequency is directly at the peak of the \mbox{F=3 $\rightarrow$ F$'$=2,3,4} absorption signal, about 9.2~GHz away from the probe frequency.

 The pump is modulated by the AOM at twice the Larmor frequency, $2f_{L}\sim$7~kHz at 1~$\mu$T.  The duty cycle and number of cycles are manually adjusted to maximize the amplitude of the signal at the start of the probe cycle.  Since the pump is shut off during the probe cycle, this is the only relevant measure of goodness for the pump parameters.  A 20\% duty cycle is typical.

To operate in 1~$\mu$T we use passive magnetic shielding to reduce both earth's field and fluctuations in it.  The shield is four nested cylinders of $\mu$-metal, a highly magnetically permeable alloy of nickel.  The outer cylinder is 40~cm long with a 15~cm radius.  Since the sensitive axis of the sensors is transverse to the long axis of the sensor (Fig.~\ref{fig:sn19}), the innermost volume contains a 3D printed saddle-wound coil designed to provide a homogeneous transverse field inside the shield. The volume inside this coil is 20~cm long with radius 3.5~cm.  The axial shielding factor has been measured~\cite{MARTIN2015} to be 1.4$\times 10^7$ with all endcaps installed, and while the transverse shielding factor has not been measured it is expected to be much higher.  Since this inner volume is quite small, the uniformity is on the order of nT/cm.  This is quite non-uniform, but provides very high AC and DC shielding factors.  All sensor tests are done in this shield.

The current for the transverse coil is provided by a custom made power supply~\cite{Ahmed24} designed and built by Shomi Ahmed at the University of Winnipeg.  The supply is designed to provide a constant 10~mA and is only adjustable via the replacement of internal components.  The supply has a monitor output that produces a directly proportional voltage by measuring across a precision 1~$\Omega$ foil resistor.  Because of the inability of this power supply to change its output, a secondary winding is layered in the same wire path to provide another nearly identical coil so that an additional nT-scale field can be applied with a second power supply, which only has to supply $\mu$A-scale current and can be battery operated to reduce noise.  This secondary supply was designed and built at TRIUMF.  This additional field allows us to measure the relative scaling of two sensors by measuring the same nT-scale field step with both sensors simultaneously.

\section{Results}\label{sec:results}

\begin{figure}
    \centering
    \includegraphics[width=0.5\textwidth]{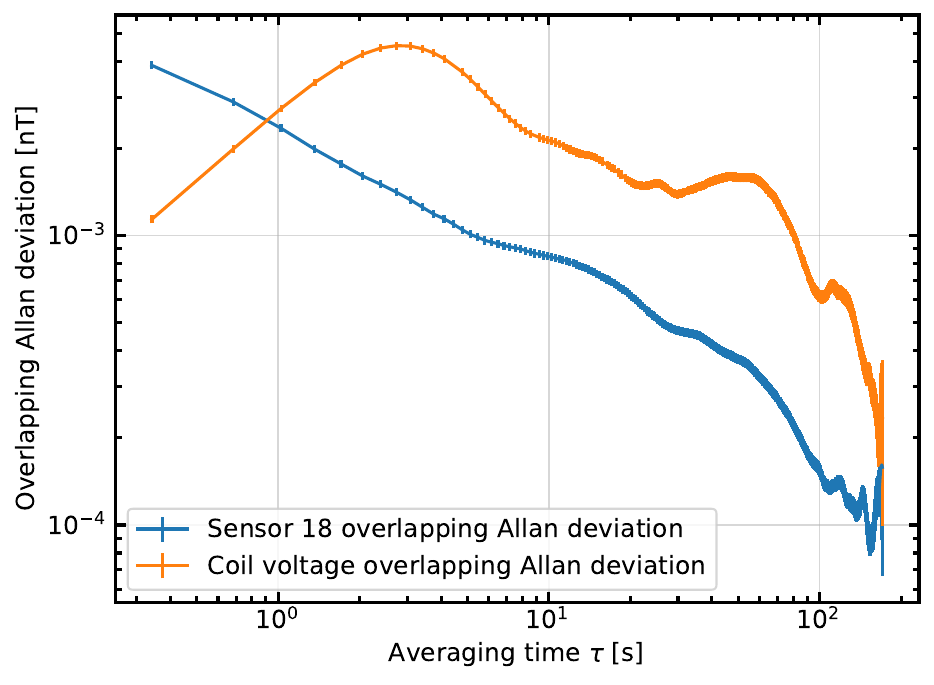}
    \caption{The Allan deviation of the magnetic field and the coil voltage measurements showing the stability of the signals at different integration times. The minimum integration time for this Allan deviation is $\sim$ \qty{350}{\ms} with a deviation of \mbox{3.9 $\pm$ 0.1}~pT, in good agreement with the Cramer-Rao lower bound.  The global minimum is \mbox{90$\pm$8}~fT at an integration time of 150 seconds}
    \label{fig:adev}
\end{figure}

Figure \ref{fig:samplemeasurement} shows a magnetic field measurement with the system.  After being split into individual FPDs the signal is filtered with a windowed-sinc FIR filter, with a pass-band from \mbox{6-8}~kHz, transitional bands of 2~kHz on each side, and $<0.04$~dB of ripple in the pass band.
The filtered signal is fit to an exponentially decaying sine wave
\begin{equation}\label{eq:fitequation}
    V(t) = Ae^{-t/T_{2}}\sin(\omega t + \Phi) + \mathrm{Offset},
\end{equation}

where $A$ is the amplitude, $T_2$ is the time constant of the decay envelope, $\omega$ is the frequency of the precession, and $\Phi$ is the phase.  $\omega$ is directly scaled to be the magnetic field strength via the gyromagnetic ratio of Cs.  While this data was taken, the coil voltage was being monitored across a precision resistor using a digital multimeter (DMM) with 8.5 digits of precision.  The voltage is scaled to a magnetic field via the ratio of the averages of the two data sets so they can be directly compared without a perfectly accurate coil constant.  The Allan deviation of the same data is shown in Figure \ref{fig:adev}.  The voltage monitor from the power supply is significantly less stable than the sensor readings for most integration times.  We attribute this to environmental effects on both the DMM itself and the connection to the monitor port.
The sensor's Allan deviation reaches a minimum of \mbox{90$\pm$8}~fT at an integration time of 150 seconds, so we can conclude that our sensors are at least this stable.  It is encouraging that we were able to get within a factor of two of the panEDM stability result~\cite{rosner2022} despite not having access to the PTB facilities.  The minimum integration time Allan deviation corresponds to the average deviation between successive points and can be taken as a measurement of sensitivity. In this case this deviation is 3.9 $\pm$ 0.1 pT.  The square root of the power spectrum is a standard measure of sensitivity, for this measurement we find 3.5~pT/$\sqrt{\mathrm{Hz}}$ at 1~Hz.

\begin{figure}
    \centering
    \includegraphics[width = 0.5\textwidth]{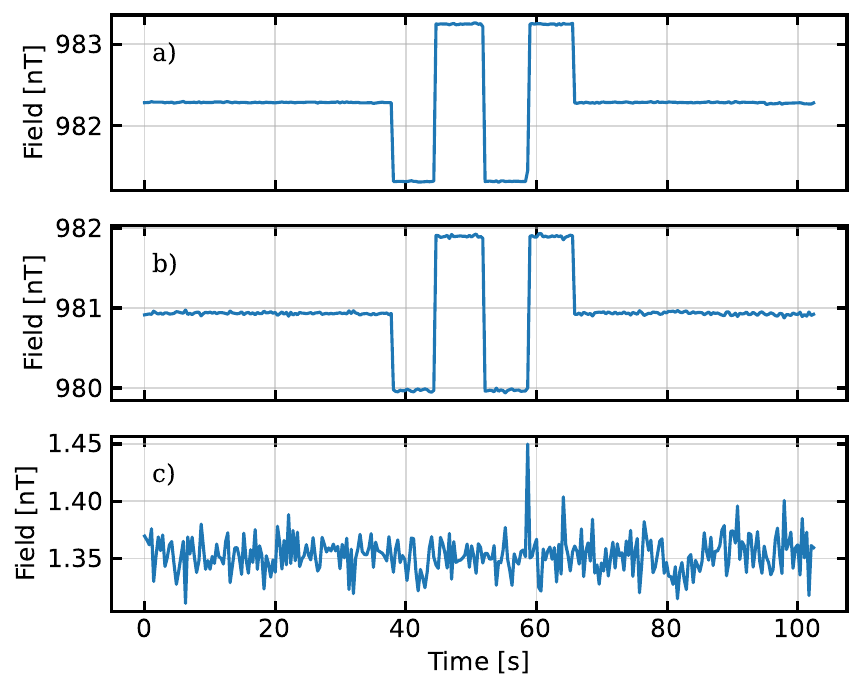}
    \caption{Magnetic field measurements of two sensors (a and b) together with
the difference between them (c) as a function of time.
The sensors are operated simultaneously, and when the field
is stepped up and down, the change in field is observed in each sensor
individually.  The sensor measurements are subtracted, as shown in (c),
and the difference remains at their combined noise floor.  Because of the relatively inhomogeneous field in the magnetic shield, the two sensors are operating at different fields and thus are not able to be perfectly optimized simultaneously given they are driven by the same optical system with a common pump cycle frequency.  The higher noise floor is the result of this.  The spike in the difference signal around 60 seconds can be associated with the changing magnetic field, and is most likely the result of the field changing partway through an FPD ringdown}
    \label{fig:2sensors}
\end{figure}

Figure \ref{fig:2sensors} shows the result of operating two such sensors simultaneously.  As they measure the field, the field is stepped up and down.  Each sensor clearly sees this deviation, but the subtraction of the fields, shown in the lower plot, remains at the noise floor of the subtracted signal.  This shows that the sensors have the same scaling to within the noise of this signal.  The overall noise in the subtracted signal is larger than the signal in Figure \ref{fig:samplemeasurement} for two reasons: the subtraction of two signals adds their variances, and the variance of each sensor is slightly higher than the minimum because each is operating at a slightly different field and thus neither is being pumped at its local Larmor frequency.

The shield is fairly inhomogeneous due to its small size, so the sensors are in slightly different fields.  Since they are serviced by a single AOM that cycles the pump light, we must compromise on what frequency at which to pump both.  The subtracted data in the lower plot in figure \ref{fig:2sensors} shows that the sensors are measuring approximately a 1.35~nT difference between the two sensor locations.

This simultaneous operation using a single laser diode with such an easily expandable system is another benefit of the commercial optics were are using.  The system as it is set up in figure \ref{fig:schematic} can operate 8 sensors with no modification whatsoever, provided sufficient magnetically shielded volume, and the TUCAN collaboration has received and successfully operated each of the full compliment of 20 such sensors.  The relatively simple addition of more fibre optic splitters will allow the full suite of sensors to be operated once the TUCAN MSR is complete.

The sensitivity limit for this system is given by the Cramer-Rao lower bound (CRLB) for a damped sinusoid, which is given by~\cite{Gemmel2010}:
\begin{equation}
    \sigma_{B} \geq \frac{\sqrt{12\cdot C}}{(A/\rho)\cdot T^{3/2}\cdot\gamma},
\end{equation}
with
\begin{equation}
    C = \frac{N^3}{12}\frac{(1-z^{2N})(1-z^2)^3}{z^2(1-z^{2N})^2 - N^2z^{2N}(1-z^2)^2},
\end{equation}
where $A$ is the amplitude of the signal, $\rho$ is the average noise density in the frequency band of interest in units of amplitude per root Hz, $T$ is the total time of the FPD signal, $\gamma$ is the gyromagnetic ratio of cesium in radians per second = 7.0196$\cdot2\pi$, $N$ is the total number of points, and $z = e^{-\frac{\Delta t}{T_2}}$ where $\Delta t$ is the sampling interval and $T_2$ is the FPD time constant as defined in equation \ref{eq:fitequation}.  $(A/\rho)$ is the signal-to-noise ratio, written as SNR.

For the data shown in Figure \ref{fig:samplemeasurement} the time constant was 40~ms.  With an average SNR of 2142 the CRLB is 3.81~pT, in good agreement with the data.

\section{Conclusion and Outlook}

We have demonstrated the simultaneous operation of multiple drift-stable, magnetically silent magnetometers with a sensitivity of 3.5~pT/$\sqrt{\mathrm{Hz}}$ at 1~Hz and a drift stability of 90~fT over 150 seconds.  This meets the requirements of the TUCAN nEDM experiment for magnetometry~\cite{klassen2020}.  Further studies demonstrating the operation of multiple sensors, measuring crosstalk, offset, and light shift are being planned for the completion of the TUCAN magnetically shielded room, currently under construction at TRIUMF.  The final magnetometric array will have 20 sensors serviced by the same probe laser, eliminating the effects of laser frequency and intensity variation on differential field measurements.

\bmhead{Acknowledgements}

The authors thank Drs. Martin Rosner and Georg Bison for helpful communications during development.  The authors also thank all of our TUCAN colleagues for support and guidance through these studies.  W. Klassen wishes to thank Dr. Derek Fujimoto for help copy-editing the final draft.

\section*{Declarations}

This research was supported by the Canada Foundation for Innovation; the Canada Research Chairs program; the Natural Sciences and Engineering Research Council of Canada (NSERC) SAPPJ-2016-00024, SAPPJ-2019-00031 and SAPPJ-2023-00029.  This work was additionally supported by JSPS KAKENHI (Grant Nos. 18H05230, 20KK0069, 20K14487 and 22H01236), JSPS Bilateral Program (Grant No. JSPSBP120239940), JST FOREST Program (Grant No. JPMJFR2237), International Joint Research Promotion Program of Osaka University, RCNP COREnet, the Yamada Science Foundation and the Murata Science Foundation.
Chris Hovde and Southwest Sciences were contracted for the design and construction of the fiberized sensors.


\bibliography{sn-bibliography}

\providecommand{\noopsort}[1]{}\providecommand{\singleletter}[1]{#1}%

\begin{thebibliography}{39}
\ifx \bisbn   \undefined \def \bisbn  #1{ISBN #1}\fi
\ifx \binits  \undefined \def \binits#1{#1}\fi
\ifx \bauthor  \undefined \def \bauthor#1{#1}\fi
\ifx \batitle  \undefined \def \batitle#1{#1}\fi
\ifx \bjtitle  \undefined \def \bjtitle#1{#1}\fi
\ifx \bvolume  \undefined \def \bvolume#1{\textbf{#1}}\fi
\ifx \byear  \undefined \def \byear#1{#1}\fi
\ifx \bissue  \undefined \def \bissue#1{#1}\fi
\ifx \bfpage  \undefined \def \bfpage#1{#1}\fi
\ifx \blpage  \undefined \def \blpage #1{#1}\fi
\ifx \burl  \undefined \def \burl#1{\textsf{#1}}\fi
\ifx \doiurl  \undefined \def \doiurl#1{\url{https://doi.org/#1}}\fi
\ifx \betal  \undefined \def \betal{\textit{et al.}}\fi
\ifx \binstitute  \undefined \def \binstitute#1{#1}\fi
\ifx \binstitutionaled  \undefined \def \binstitutionaled#1{#1}\fi
\ifx \bctitle  \undefined \def \bctitle#1{#1}\fi
\ifx \beditor  \undefined \def \beditor#1{#1}\fi
\ifx \bpublisher  \undefined \def \bpublisher#1{#1}\fi
\ifx \bbtitle  \undefined \def \bbtitle#1{#1}\fi
\ifx \bedition  \undefined \def \bedition#1{#1}\fi
\ifx \bseriesno  \undefined \def \bseriesno#1{#1}\fi
\ifx \blocation  \undefined \def \blocation#1{#1}\fi
\ifx \bsertitle  \undefined \def \bsertitle#1{#1}\fi
\ifx \bsnm \undefined \def \bsnm#1{#1}\fi
\ifx \bsuffix \undefined \def \bsuffix#1{#1}\fi
\ifx \bparticle \undefined \def \bparticle#1{#1}\fi
\ifx \barticle \undefined \def \barticle#1{#1}\fi
\bibcommenthead
\ifx \bconfdate \undefined \def \bconfdate #1{#1}\fi
\ifx \botherref \undefined \def \botherref #1{#1}\fi
\ifx \url \undefined \def \url#1{\textsf{#1}}\fi
\ifx \bchapter \undefined \def \bchapter#1{#1}\fi
\ifx \bbook \undefined \def \bbook#1{#1}\fi
\ifx \bcomment \undefined \def \bcomment#1{#1}\fi
\ifx \oauthor \undefined \def \oauthor#1{#1}\fi
\ifx \citeauthoryear \undefined \def \citeauthoryear#1{#1}\fi
\ifx \endbibitem  \undefined \def \endbibitem {}\fi
\ifx \bconflocation  \undefined \def \bconflocation#1{#1}\fi
\ifx \arxivurl  \undefined \def \arxivurl#1{\textsf{#1}}\fi
\csname PreBibitemsHook\endcsname

\bibitem[\protect\citeauthoryear{{Wurm, David} et~al.}{2019}]{panEDM2019}
\begin{barticle}
\bauthor{\bsnm{{Wurm, David}}},
\bauthor{\bsnm{{Beck, Douglas H.}}},
\bauthor{\bsnm{{Chupp, Tim}}},
\bauthor{\bsnm{{Degenkolb, Skyler}}},
\bauthor{\bsnm{{Fierlinger, Katharina}}},
\bauthor{\bsnm{{Fierlinger, Peter}}},
\bauthor{\bsnm{{Filter, Hanno}}},
\bauthor{\bsnm{{Ivanov, Sergey}}},
\bauthor{\bsnm{{Klau, Christopher}}},
\bauthor{\bsnm{{Kreuz, Michael}}},
\bauthor{\bsnm{{Leli\`evre-Berna, Eddy}}},
\bauthor{\bsnm{{Lins, Tobias}}},
\bauthor{\bsnm{{Meichelb\"ock, Joachim}}},
\bauthor{\bsnm{{Neulinger, Thomas}}},
\bauthor{\bsnm{{Paddock, Robert}}},
\bauthor{\bsnm{{R\"ohrer, Florian}}},
\bauthor{\bsnm{{Rosner, Martin}}},
\bauthor{\bsnm{{Serebrov, Anatolii P.}}},
\bauthor{\bsnm{{Singh, Jaideep Taggart}}},
\bauthor{\bsnm{{Stoepler, Rainer}}},
\bauthor{\bsnm{{Stuiber, Stefan}}},
\bauthor{\bsnm{{Sturm, Michael}}},
\bauthor{\bsnm{{Taubenheim, Bernd}}},
\bauthor{\bsnm{{Tonon, Xavier}}},
\bauthor{\bsnm{{Tucker, Mark}}},
\bauthor{\bsnm{{Grinten, Maurits van der}}},
\bauthor{\bsnm{{Zimmer, Oliver}}}:
\batitle{{The PanEDM neutron electric dipole moment experiment at the ILL}}.
\bjtitle{EPJ Web Conf.}
\bvolume{219},
\bfpage{02006}
(\byear{2019})
\doiurl{10.1051/epjconf/201921902006}
\end{barticle}
\endbibitem

\bibitem[\protect\citeauthoryear{{Abel, C.} et~al.}{2019}]{n2EDM2019}
\begin{barticle}
\bauthor{\bsnm{{Abel, C.}}},
\bauthor{\bsnm{{Ayres, N. J.}}},
\bauthor{\bsnm{{Ban, G.}}},
\bauthor{\bsnm{{Bison, G.}}},
\bauthor{\bsnm{{Bodek, K.}}},
\bauthor{\bsnm{{Bondar, V.}}},
\bauthor{\bsnm{{Chanel, E.}}},
\bauthor{\bsnm{{Chiu, P.-J.}}},
\bauthor{\bsnm{{Clement, B.}}},
\bauthor{\bsnm{{Crawford, C.}}},
\bauthor{\bsnm{{Daum, M.}}},
\bauthor{\bsnm{{Emmenegger, S.}}},
\bauthor{\bsnm{{Flaux, P.}}},
\bauthor{\bsnm{{Ferraris-Bouchez, L.}}},
\bauthor{\bsnm{{Griffith, W.C.}}},
\bauthor{\bsnm{{Gruji\'{c}, Z.D.}}},
\bauthor{\bsnm{{Harris, P.G.}}},
\bauthor{\bsnm{{Heil, W.}}},
\bauthor{\bsnm{{Hild, N.}}},
\bauthor{\bsnm{{Kirch, K.}}},
\bauthor{\bsnm{{Koss, P.A.}}},
\bauthor{\bsnm{{Kozela, A.}}},
\bauthor{\bsnm{{Krempel, J.}}},
\bauthor{\bsnm{{Lauss, B.}}},
\bauthor{\bsnm{{Lefort, T.}}},
\bauthor{\bsnm{{Lemi\`ere, Y.}}},
\bauthor{\bsnm{{Leredde, A.}}},
\bauthor{\bsnm{{Mohanmurthy, P.}}},
\bauthor{\bsnm{{Naviliat-Cuncic, O.}}},
\bauthor{\bsnm{{Pais, D.}}},
\bauthor{\bsnm{{Piegsa, F.M.}}},
\bauthor{\bsnm{{Pignol, G.}}},
\bauthor{\bsnm{{Rawlik, M.}}},
\bauthor{\bsnm{{Rebreyend, D.}}},
\bauthor{\bsnm{{Ries, D.}}},
\bauthor{\bsnm{{Roccia, S.}}},
\bauthor{\bsnm{{Ross, K.}}},
\bauthor{\bsnm{{Rozpedzik, D.}}},
\bauthor{\bsnm{{Schmidt-Wellenburg, P.}}},
\bauthor{\bsnm{{Schnabel, A.}}},
\bauthor{\bsnm{{Severijns, N.}}},
\bauthor{\bsnm{{Thorne, J.}}},
\bauthor{\bsnm{{Virot, R.}}},
\bauthor{\bsnm{{Voigt, J.}}},
\bauthor{\bsnm{{Weis, A.}}},
\bauthor{\bsnm{{Wursten, E.}}},
\bauthor{\bsnm{{Zejma, J.}}},
\bauthor{\bsnm{{Zsigmond, G.}}}:
\batitle{{The n2EDM experiment at the Paul Scherrer Institute}}.
\bjtitle{EPJ Web Conf.}
\bvolume{219},
\bfpage{02002}
(\byear{2019})
\doiurl{10.1051/epjconf/201921902002}
\end{barticle}
\endbibitem

\bibitem[\protect\citeauthoryear{Ito et~al.}{2018}]{lanl2018}
\begin{barticle}
\bauthor{\bsnm{Ito}, \binits{T.M.}},
\bauthor{\bsnm{Adamek}, \binits{E.R.}},
\bauthor{\bsnm{Callahan}, \binits{N.B.}},
\bauthor{\bsnm{Choi}, \binits{J.H.}},
\bauthor{\bsnm{Clayton}, \binits{S.M.}},
\bauthor{\bsnm{Cude-Woods}, \binits{C.}},
\bauthor{\bsnm{Currie}, \binits{S.}},
\bauthor{\bsnm{Ding}, \binits{X.}},
\bauthor{\bsnm{Fellers}, \binits{D.E.}},
\bauthor{\bsnm{Geltenbort}, \binits{P.}},
\bauthor{\bsnm{Lamoreaux}, \binits{S.K.}},
\bauthor{\bsnm{Liu}, \binits{C.-Y.}},
\bauthor{\bsnm{MacDonald}, \binits{S.}},
\bauthor{\bsnm{Makela}, \binits{M.}},
\bauthor{\bsnm{Morris}, \binits{C.L.}},
\bauthor{\bsnm{Pattie}, \binits{R.W.}},
\bauthor{\bsnm{Ramsey}, \binits{J.C.}},
\bauthor{\bsnm{Salvat}, \binits{D.J.}},
\bauthor{\bsnm{Saunders}, \binits{A.}},
\bauthor{\bsnm{Sharapov}, \binits{E.I.}},
\bauthor{\bsnm{Sjue}, \binits{S.}},
\bauthor{\bsnm{Sprow}, \binits{A.P.}},
\bauthor{\bsnm{Tang}, \binits{Z.}},
\bauthor{\bsnm{Weaver}, \binits{H.L.}},
\bauthor{\bsnm{Wei}, \binits{W.}},
\bauthor{\bsnm{Young}, \binits{A.R.}}:
\batitle{{Performance of the upgraded ultracold neutron source at Los Alamos National Laboratory and its implication for a possible neutron electric dipole moment experiment}}.
\bjtitle{Phys. Rev. C}
\bvolume{97},
\bfpage{012501}
(\byear{2018})
\doiurl{10.1103/PhysRevC.97.012501}
\end{barticle}
\endbibitem

\bibitem[\protect\citeauthoryear{Serebrov et~al.}{2015}]{pnpi_ILL2015}
\begin{barticle}
\bauthor{\bsnm{Serebrov}, \binits{A.P.}},
\bauthor{\bsnm{Kolomenskiy}, \binits{E.A.}},
\bauthor{\bsnm{Pirozhkov}, \binits{A.N.}},
\bauthor{\bsnm{Krasnoschekova}, \binits{I.A.}},
\bauthor{\bsnm{Vassiljev}, \binits{A.V.}},
\bauthor{\bsnm{Polyushkin}, \binits{A.O.}},
\bauthor{\bsnm{Lasakov}, \binits{M.S.}},
\bauthor{\bsnm{Murashkin}, \binits{A.N.}},
\bauthor{\bsnm{Solovey}, \binits{V.A.}},
\bauthor{\bsnm{Fomin}, \binits{A.K.}},
\bauthor{\bsnm{Shoka}, \binits{I.V.}},
\bauthor{\bsnm{Zherebtsov}, \binits{O.M.}},
\bauthor{\bsnm{Geltenbort}, \binits{P.}},
\bauthor{\bsnm{Ivanov}, \binits{S.N.}},
\bauthor{\bsnm{Zimmer}, \binits{O.}},
\bauthor{\bsnm{Alexandrov}, \binits{E.B.}},
\bauthor{\bsnm{Dmitriev}, \binits{S.P.}},
\bauthor{\bsnm{Dovator}, \binits{N.A.}}:
\batitle{{New search for the neutron electric dipole moment with ultracold neutrons at ILL}}.
\bjtitle{Phys. Rev. C}
\bvolume{92},
\bfpage{055501}
(\byear{2015})
\doiurl{10.1103/PhysRevC.92.055501}
\end{barticle}
\endbibitem

\bibitem[\protect\citeauthoryear{Ahmed et~al.}{2019}]{sns_2019}
\begin{barticle}
\bauthor{\bsnm{Ahmed}, \binits{M.W.}},
\bauthor{\bsnm{Alarcon}, \binits{R.}},
\bauthor{\bsnm{Aleksandrova}, \binits{A.}},
\bauthor{\bsnm{Baeßler}, \binits{S.}},
\bauthor{\bsnm{Barron-Palos}, \binits{L.}},
\bauthor{\bsnm{Bartoszek}, \binits{L.M.}},
\bauthor{\bsnm{Beck}, \binits{D.H.}},
\bauthor{\bsnm{Behzadipour}, \binits{M.}},
\bauthor{\bsnm{Berkutov}, \binits{I.}},
\bauthor{\bsnm{Bessuille}, \binits{J.}},
\bauthor{\bsnm{Blatnik}, \binits{M.}},
\bauthor{\bsnm{Broering}, \binits{M.}},
\bauthor{\bsnm{Broussard}, \binits{L.J.}},
\bauthor{\bsnm{Busch}, \binits{M.}},
\bauthor{\bsnm{Carr}, \binits{R.}},
\bauthor{\bsnm{Cianciolo}, \binits{V.}},
\bauthor{\bsnm{Clayton}, \binits{S.M.}},
\bauthor{\bsnm{Cooper}, \binits{M.D.}},
\bauthor{\bsnm{Crawford}, \binits{C.}},
\bauthor{\bsnm{Currie}, \binits{S.A.}},
\bauthor{\bsnm{Daurer}, \binits{C.}},
\bauthor{\bsnm{Dipert}, \binits{R.}},
\bauthor{\bsnm{Dow}, \binits{K.}},
\bauthor{\bsnm{Dutta}, \binits{D.}},
\bauthor{\bsnm{Efremenko}, \binits{Y.}},
\bauthor{\bsnm{Erickson}, \binits{C.B.}},
\bauthor{\bsnm{Filippone}, \binits{B.W.}},
\bauthor{\bsnm{Fomin}, \binits{N.}},
\bauthor{\bsnm{Gao}, \binits{H.}},
\bauthor{\bsnm{Golub}, \binits{R.}},
\bauthor{\bsnm{Gould}, \binits{C.R.}},
\bauthor{\bsnm{Greene}, \binits{G.}},
\bauthor{\bsnm{Haase}, \binits{D.G.}},
\bauthor{\bsnm{Hasell}, \binits{D.}},
\bauthor{\bsnm{Hawari}, \binits{A.I.}},
\bauthor{\bsnm{Hayden}, \binits{M.E.}},
\bauthor{\bsnm{Holley}, \binits{A.}},
\bauthor{\bsnm{Holt}, \binits{R.J.}},
\bauthor{\bsnm{Huffman}, \binits{P.R.}},
\bauthor{\bsnm{Ihloff}, \binits{E.}},
\bauthor{\bsnm{Imam}, \binits{S.K.}},
\bauthor{\bsnm{Ito}, \binits{T.M.}},
\bauthor{\bsnm{Karcz}, \binits{M.}},
\bauthor{\bsnm{Kelsey}, \binits{J.}},
\bauthor{\bsnm{Kendellen}, \binits{D.P.}},
\bauthor{\bsnm{Kim}, \binits{Y.J.}},
\bauthor{\bsnm{Korobkina}, \binits{E.}},
\bauthor{\bsnm{Korsch}, \binits{W.}},
\bauthor{\bsnm{Lamoreaux}, \binits{S.K.}},
\bauthor{\bsnm{Leggett}, \binits{E.}},
\bauthor{\bsnm{Leung}, \binits{K.K.H.}},
\bauthor{\bsnm{Lipman}, \binits{A.}},
\bauthor{\bsnm{Liu}, \binits{C.Y.}},
\bauthor{\bsnm{Long}, \binits{J.}},
\bauthor{\bsnm{MacDonald}, \binits{S.W.T.}},
\bauthor{\bsnm{Makela}, \binits{M.}},
\bauthor{\bsnm{Matlashov}, \binits{A.}},
\bauthor{\bsnm{Maxwell}, \binits{J.D.}},
\bauthor{\bsnm{Mendenhall}, \binits{M.}},
\bauthor{\bsnm{Meyer}, \binits{H.O.}},
\bauthor{\bsnm{Milner}, \binits{R.G.}},
\bauthor{\bsnm{Mueller}, \binits{P.E.}},
\bauthor{\bsnm{Nouri}, \binits{N.}},
\bauthor{\bsnm{O'Shaughnessy}, \binits{C.M.}},
\bauthor{\bsnm{Osthelder}, \binits{C.}},
\bauthor{\bsnm{Peng}, \binits{J.C.}},
\bauthor{\bsnm{Penttila}, \binits{S.I.}},
\bauthor{\bsnm{Phan}, \binits{N.S.}},
\bauthor{\bsnm{Plaster}, \binits{B.}},
\bauthor{\bsnm{Ramsey}, \binits{J.C.}},
\bauthor{\bsnm{Rao}, \binits{T.M.}},
\bauthor{\bsnm{Redwine}, \binits{R.P.}},
\bauthor{\bsnm{Reid}, \binits{A.}},
\bauthor{\bsnm{Saftah}, \binits{A.}},
\bauthor{\bsnm{Seidel}, \binits{G.M.}},
\bauthor{\bsnm{Silvera}, \binits{I.}},
\bauthor{\bsnm{Slutsky}, \binits{S.}},
\bauthor{\bsnm{Smith}, \binits{E.}},
\bauthor{\bsnm{Snow}, \binits{W.M.}},
\bauthor{\bsnm{Sondheim}, \binits{W.}},
\bauthor{\bsnm{Sosothikul}, \binits{S.}},
\bauthor{\bsnm{Stanislaus}, \binits{T.D.S.}},
\bauthor{\bsnm{Sun}, \binits{X.}},
\bauthor{\bsnm{Swank}, \binits{C.M.}},
\bauthor{\bsnm{Tang}, \binits{Z.}},
\bauthor{\bsnm{Dinani}, \binits{R.T.}},
\bauthor{\bsnm{Tsentalovich}, \binits{E.}},
\bauthor{\bsnm{Vidal}, \binits{C.}},
\bauthor{\bsnm{Wei}, \binits{W.}},
\bauthor{\bsnm{White}, \binits{C.R.}},
\bauthor{\bsnm{Williamson}, \binits{S.E.}},
\bauthor{\bsnm{Yang}, \binits{L.}},
\bauthor{\bsnm{Yao}, \binits{W.}},
\bauthor{\bsnm{Young}, \binits{A.R.}}:
\batitle{{A new cryogenic apparatus to search for the neutron electric dipole moment}}.
\bjtitle{JINST}
\bvolume{14}(\bissue{11}),
\bfpage{11017}
(\byear{2019})
\doiurl{10.1088/1748-0221/14/11/P11017}
\end{barticle}
\endbibitem

\bibitem[\protect\citeauthoryear{Itoh et~al.}{2018}]{JPARC_2018}
\begin{barticle}
\bauthor{\bsnm{Itoh}, \binits{S.}},
\bauthor{\bsnm{Nakaji}, \binits{M.}},
\bauthor{\bsnm{Uchida}, \binits{Y.}},
\bauthor{\bsnm{Kitaguchi}, \binits{M.}},
\bauthor{\bsnm{Shimizu}, \binits{H.M.}}:
\batitle{{Pendellösung interferometry by using pulsed neutrons}}.
\bjtitle{NIM-A}
\bvolume{908},
\bfpage{78}--\blpage{81}
(\byear{2018})
\doiurl{10.1016/j.nima.2018.08.043}
\end{barticle}
\endbibitem

\bibitem[\protect\citeauthoryear{Carena et~al.}{2019}]{carena_2019}
\begin{barticle}
\bauthor{\bsnm{Carena}, \binits{M.}},
\bauthor{\bsnm{Liu}, \binits{D.}},
\bauthor{\bsnm{Liu}, \binits{J.}},
\bauthor{\bsnm{Shah}, \binits{N.R.}},
\bauthor{\bsnm{Wagner}, \binits{C.E.M.}},
\bauthor{\bsnm{Wang}, \binits{X.-P.}}:
\batitle{{$\ensuremath{\nu}$ solution to the strong $CP$ problem}}.
\bjtitle{Phys. Rev. D}
\bvolume{100},
\bfpage{094018}
(\byear{2019})
\doiurl{10.1103/PhysRevD.100.094018}
\end{barticle}
\endbibitem

\bibitem[\protect\citeauthoryear{Mimura et~al.}{2019}]{Mimura_2019}
\begin{barticle}
\bauthor{\bsnm{Mimura}, \binits{Y.}},
\bauthor{\bsnm{Mohapatra}, \binits{R.N.}},
\bauthor{\bsnm{Severson}, \binits{M.}}:
\batitle{{Grand unified parity solution to the strong $CP$ problem}}.
\bjtitle{Phys. Rev. D}
\bvolume{99},
\bfpage{115025}
(\byear{2019})
\doiurl{10.1103/PhysRevD.99.115025}
\end{barticle}
\endbibitem

\bibitem[\protect\citeauthoryear{Morrissey and Ramsey-Musolf}{2012}]{Morrissey_2012}
\begin{barticle}
\bauthor{\bsnm{Morrissey}, \binits{D.E.}},
\bauthor{\bsnm{Ramsey-Musolf}, \binits{M.J.}}:
\batitle{{Electroweak baryogenesis}}.
\bjtitle{NJP}
\bvolume{14}(\bissue{12}),
\bfpage{125003}
(\byear{2012})
\doiurl{10.1088/1367-2630/14/12/125003}
\end{barticle}
\endbibitem

\bibitem[\protect\citeauthoryear{Bell et~al.}{2019}]{nfbell_2019}
\begin{barticle}
\bauthor{\bsnm{Bell}, \binits{N.F.}},
\bauthor{\bsnm{Corbett}, \binits{T.}},
\bauthor{\bsnm{Nee}, \binits{M.}},
\bauthor{\bsnm{Ramsey-Musolf}, \binits{M.J.}}:
\batitle{{Electric dipole moments from postsphaleron baryogenesis}}.
\bjtitle{Phys. Rev. D}
\bvolume{99},
\bfpage{015034}
(\byear{2019})
\doiurl{10.1103/PhysRevD.99.015034}
\end{barticle}
\endbibitem

\bibitem[\protect\citeauthoryear{Cirigliano et~al.}{2019}]{cirigliano_2019}
\begin{barticle}
\bauthor{\bsnm{Cirigliano}, \binits{V.}},
\bauthor{\bsnm{Crivellin}, \binits{A.}},
\bauthor{\bsnm{Dekens}, \binits{W.}},
\bauthor{\bsnm{Vries}, \binits{J.}},
\bauthor{\bsnm{Hoferichter}, \binits{M.}},
\bauthor{\bsnm{Mereghetti}, \binits{E.}}:
\batitle{{$CP$ Violation in Higgs-Gauge Interactions: From Tabletop Experiments to the LHC}}.
\bjtitle{Phys. Rev. Lett.}
\bvolume{123},
\bfpage{051801}
(\byear{2019})
\doiurl{10.1103/PhysRevLett.123.051801}
\end{barticle}
\endbibitem

\bibitem[\protect\citeauthoryear{Crivellin and Saturnino}{2019}]{Crivellin_Saturnino_2019}
\begin{barticle}
\bauthor{\bsnm{Crivellin}, \binits{A.}},
\bauthor{\bsnm{Saturnino}, \binits{F.}}:
\batitle{{Correlating tauonic $B$ decays with the neutron electric dipole moment via a scalar leptoquark}}.
\bjtitle{Phys. Rev. D}
\bvolume{100},
\bfpage{115014}
(\byear{2019})
\doiurl{10.1103/PhysRevD.100.115014}
\end{barticle}
\endbibitem

\bibitem[\protect\citeauthoryear{Andreev et~al.}{2018}]{acme_2018}
\begin{barticle}
\bauthor{\bsnm{Andreev}, \binits{V.}},
\bauthor{\bsnm{Ang}, \binits{D.G.}},
\bauthor{\bsnm{DeMille}, \binits{D.}},
\bauthor{\bsnm{Doyle}, \binits{J.M.}},
\bauthor{\bsnm{Gabrielse}, \binits{G.}},
\bauthor{\bsnm{Haefner}, \binits{J.}},
\bauthor{\bsnm{Hutzler}, \binits{N.R.}},
\bauthor{\bsnm{Lasner}, \binits{Z.}},
\bauthor{\bsnm{Meisenhelder}, \binits{C.}},
\bauthor{\bsnm{O'Leary}, \binits{B.R.}},
\bauthor{\bsnm{Panda}, \binits{C.D.}},
\bauthor{\bsnm{West}, \binits{A.D.}},
\bauthor{\bsnm{West}, \binits{E.P.}},
\bauthor{\bsnm{Wu}, \binits{X.}},
\bauthor{\bsnm{Collaboration}, \binits{A.C.M.E.}}:
\batitle{{Improved limit on the electric dipole moment of the electron}}.
\bjtitle{Nature}
\bvolume{562}(\bissue{7727}),
\bfpage{355}--\blpage{360}
(\byear{2018})
\doiurl{10.1038/s41586-018-0599-8}
\end{barticle}
\endbibitem

\bibitem[\protect\citeauthoryear{Abel et~al.}{2020}]{psi2020}
\begin{barticle}
\bauthor{\bsnm{Abel}, \binits{C.}},
\bauthor{\bsnm{Afach}, \binits{S.}},
\bauthor{\bsnm{Ayres}, \binits{N.J.}},
\bauthor{\bsnm{Baker}, \binits{C.A.}},
\bauthor{\bsnm{Ban}, \binits{G.}},
\bauthor{\bsnm{Bison}, \binits{G.}},
\bauthor{\bsnm{Bodek}, \binits{K.}},
\bauthor{\bsnm{Bondar}, \binits{V.}},
\bauthor{\bsnm{Burghoff}, \binits{M.}},
\bauthor{\bsnm{Chanel}, \binits{E.}},
\bauthor{\bsnm{Chowdhuri}, \binits{Z.}},
\bauthor{\bsnm{Chiu}, \binits{P.-J.}},
\bauthor{\bsnm{Clement}, \binits{B.}},
\bauthor{\bsnm{Crawford}, \binits{C.B.}},
\bauthor{\bsnm{Daum}, \binits{M.}},
\bauthor{\bsnm{Emmenegger}, \binits{S.}},
\bauthor{\bsnm{Ferraris-Bouchez}, \binits{L.}},
\bauthor{\bsnm{Fertl}, \binits{M.}},
\bauthor{\bsnm{Flaux}, \binits{P.}},
\bauthor{\bsnm{Franke}, \binits{B.}},
\bauthor{\bsnm{Fratangelo}, \binits{A.}},
\bauthor{\bsnm{Geltenbort}, \binits{P.}},
\bauthor{\bsnm{Green}, \binits{K.}},
\bauthor{\bsnm{Griffith}, \binits{W.C.}},
\bauthor{\bsnm{Grinten}, \binits{M.}},
\bauthor{\bsnm{Gruji\'{c}}, \binits{Z.D.}},
\bauthor{\bsnm{Harris}, \binits{P.G.}},
\bauthor{\bsnm{Hayen}, \binits{L.}},
\bauthor{\bsnm{Heil}, \binits{W.}},
\bauthor{\bsnm{Henneck}, \binits{R.}},
\bauthor{\bsnm{H\'elaine}, \binits{V.}},
\bauthor{\bsnm{Hild}, \binits{N.}},
\bauthor{\bsnm{Hodge}, \binits{Z.}},
\bauthor{\bsnm{Horras}, \binits{M.}},
\bauthor{\bsnm{Iaydjiev}, \binits{P.}},
\bauthor{\bsnm{Ivanov}, \binits{S.N.}},
\bauthor{\bsnm{Kasprzak}, \binits{M.}},
\bauthor{\bsnm{Kermaidic}, \binits{Y.}},
\bauthor{\bsnm{Kirch}, \binits{K.}},
\bauthor{\bsnm{Knecht}, \binits{A.}},
\bauthor{\bsnm{Knowles}, \binits{P.}},
\bauthor{\bsnm{Koch}, \binits{H.-C.}},
\bauthor{\bsnm{Koss}, \binits{P.A.}},
\bauthor{\bsnm{Komposch}, \binits{S.}},
\bauthor{\bsnm{Kozela}, \binits{A.}},
\bauthor{\bsnm{Kraft}, \binits{A.}},
\bauthor{\bsnm{Krempel}, \binits{J.}},
\bauthor{\bsnm{Ku\'{z}niak}, \binits{M.}},
\bauthor{\bsnm{Lauss}, \binits{B.}},
\bauthor{\bsnm{Lefort}, \binits{T.}},
\bauthor{\bsnm{Lemi\`ere}, \binits{Y.}},
\bauthor{\bsnm{Leredde}, \binits{A.}},
\bauthor{\bsnm{Mohanmurthy}, \binits{P.}},
\bauthor{\bsnm{Mtchedlishvili}, \binits{A.}},
\bauthor{\bsnm{Musgrave}, \binits{M.}},
\bauthor{\bsnm{Naviliat-Cuncic}, \binits{O.}},
\bauthor{\bsnm{Pais}, \binits{D.}},
\bauthor{\bsnm{Piegsa}, \binits{F.M.}},
\bauthor{\bsnm{Pierre}, \binits{E.}},
\bauthor{\bsnm{Pignol}, \binits{G.}},
\bauthor{\bsnm{Plonka-Spehr}, \binits{C.}},
\bauthor{\bsnm{Prashanth}, \binits{P.N.}},
\bauthor{\bsnm{Qu\'em\'ener}, \binits{G.}},
\bauthor{\bsnm{Rawlik}, \binits{M.}},
\bauthor{\bsnm{Rebreyend}, \binits{D.}},
\bauthor{\bsnm{Rien\"acker}, \binits{I.}},
\bauthor{\bsnm{Ries}, \binits{D.}},
\bauthor{\bsnm{Roccia}, \binits{S.}},
\bauthor{\bsnm{Rogel}, \binits{G.}},
\bauthor{\bsnm{Rozpedzik}, \binits{D.}},
\bauthor{\bsnm{Schnabel}, \binits{A.}},
\bauthor{\bsnm{Schmidt-Wellenburg}, \binits{P.}},
\bauthor{\bsnm{Severijns}, \binits{N.}},
\bauthor{\bsnm{Shiers}, \binits{D.}},
\bauthor{\bsnm{Tavakoli~Dinani}, \binits{R.}},
\bauthor{\bsnm{Thorne}, \binits{J.A.}},
\bauthor{\bsnm{Virot}, \binits{R.}},
\bauthor{\bsnm{Voigt}, \binits{J.}},
\bauthor{\bsnm{Weis}, \binits{A.}},
\bauthor{\bsnm{Wursten}, \binits{E.}},
\bauthor{\bsnm{Wyszynski}, \binits{G.}},
\bauthor{\bsnm{Zejma}, \binits{J.}},
\bauthor{\bsnm{Zenner}, \binits{J.}},
\bauthor{\bsnm{Zsigmond}, \binits{G.}}:
\batitle{{Measurement of the Permanent Electric Dipole Moment of the Neutron}}.
\bjtitle{Phys. Rev. Lett.}
\bvolume{124},
\bfpage{081803}
(\byear{2020})
\doiurl{10.1103/PhysRevLett.124.081803}
\end{barticle}
\endbibitem

\bibitem[\protect\citeauthoryear{Matsumiya et~al.}{}]{tucanTRIUMF}
\begin{botherref}
\oauthor{\bsnm{Matsumiya}, \binits{R.}},
\oauthor{\bsnm{Akatsuka}, \binits{H.}},
\oauthor{\bsnm{Bidinosti}, \binits{C.P.}},
\oauthor{\bsnm{Davis}, \binits{C.A.}},
\oauthor{\bsnm{Franke}, \binits{B.}},
\oauthor{\bsnm{Fujimoto}, \binits{D.}},
\oauthor{\bsnm{Gericke}, \binits{M.T.W.}},
\oauthor{\bsnm{Giampa}, \binits{P.}},
\oauthor{\bsnm{Golub}, \binits{R.}},
\oauthor{\bsnm{Hansen-Romu}, \binits{S.}},
\oauthor{\bsnm{Hatanaka}, \binits{K.}},
\oauthor{\bsnm{Hayamizu}, \binits{T.}},
\oauthor{\bsnm{Higuchi}, \binits{T.}},
\oauthor{\bsnm{Ichikawa}, \binits{G.}},
\oauthor{\bsnm{Imajo}, \binits{S.}},
\oauthor{\bsnm{Jamieson}, \binits{B.}},
\oauthor{\bsnm{Kawasaki}, \binits{S.}},
\oauthor{\bsnm{Kitaguchi}, \binits{M.}},
\oauthor{\bsnm{Klassen}, \binits{W.}},
\oauthor{\bsnm{Klemets}, \binits{E.}},
\oauthor{\bsnm{Konaka}, \binits{A.}},
\oauthor{\bsnm{Korkmaz}, \binits{E.}},
\oauthor{\bsnm{Korobkina}, \binits{E.}},
\oauthor{\bsnm{Kuchler}, \binits{F.}},
\oauthor{\bsnm{Lavvaf}, \binits{M.}},
\oauthor{\bsnm{Lee}, \binits{L.}},
\oauthor{\bsnm{Lindner}, \binits{T.}},
\oauthor{\bsnm{Madison}, \binits{K.W.}},
\oauthor{\bsnm{Makida}, \binits{Y.}},
\oauthor{\bsnm{Mammei}, \binits{R.}},
\oauthor{\bsnm{Mammei}, \binits{J.}},
\oauthor{\bsnm{Martin}, \binits{J.W.}},
\oauthor{\bsnm{McCrea}, \binits{M.}},
\oauthor{\bsnm{Miller}, \binits{E.}},
\oauthor{\bsnm{Mishima}, \binits{K.}},
\oauthor{\bsnm{Momose}, \binits{T.}},
\oauthor{\bsnm{Okamura}, \binits{T.}},
\oauthor{\bsnm{Ong}, \binits{H.J.}},
\oauthor{\bsnm{Picker}, \binits{R.}},
\oauthor{\bsnm{Ramsay}, \binits{W.D.}},
\oauthor{\bsnm{Schreyer}, \binits{W.}},
\oauthor{\bsnm{Shimizu}, \binits{H.M.}},
\oauthor{\bsnm{Sidhu}, \binits{S.}},
\oauthor{\bsnm{Stargardter}, \binits{S.}},
\oauthor{\bsnm{Tanihata}, \binits{I.}},
\oauthor{\bsnm{Vanbergen}, \binits{S.}},
\oauthor{\bsnm{Oers}, \binits{W.T.H.}},
\oauthor{\bsnm{Watanabe}, \binits{Y.}}:
{The Precision nEDM Measurement with UltraCold Neutrons at TRIUMF}.
JPS Conf. Proc.
\doiurl{10.7566/JPSCP.37.020701}
\end{botherref}
\endbibitem

\bibitem[\protect\citeauthoryear{Ahmed et~al.}{2019a}]{TUCAN_firstucn}
\begin{barticle}
\bauthor{\bsnm{Ahmed}, \binits{S.}},
\bauthor{\bsnm{Altiere}, \binits{E.}},
\bauthor{\bsnm{Andalib}, \binits{T.}},
\bauthor{\bsnm{Bell}, \binits{B.}},
\bauthor{\bsnm{Bidinosti}, \binits{C.P.}},
\bauthor{\bsnm{Cudmore}, \binits{E.}},
\bauthor{\bsnm{Das}, \binits{M.}},
\bauthor{\bsnm{Davis}, \binits{C.A.}},
\bauthor{\bsnm{Franke}, \binits{B.}},
\bauthor{\bsnm{Gericke}, \binits{M.}},
\bauthor{\bsnm{Giampa}, \binits{P.}},
\bauthor{\bsnm{Gnyp}, \binits{P.}},
\bauthor{\bsnm{Hansen-Romu}, \binits{S.}},
\bauthor{\bsnm{Hatanaka}, \binits{K.}},
\bauthor{\bsnm{Hayamizu}, \binits{T.}},
\bauthor{\bsnm{Jamieson}, \binits{B.}},
\bauthor{\bsnm{Jones}, \binits{D.}},
\bauthor{\bsnm{Kawasaki}, \binits{S.}},
\bauthor{\bsnm{Kikawa}, \binits{T.}},
\bauthor{\bsnm{Kitaguchi}, \binits{M.}},
\bauthor{\bsnm{Klassen}, \binits{W.}},
\bauthor{\bsnm{Konaka}, \binits{A.}},
\bauthor{\bsnm{Korkmaz}, \binits{E.}},
\bauthor{\bsnm{Kuchler}, \binits{F.}},
\bauthor{\bsnm{Lang}, \binits{M.}},
\bauthor{\bsnm{Lee}, \binits{L.}},
\bauthor{\bsnm{Lindner}, \binits{T.}},
\bauthor{\bsnm{Madison}, \binits{K.W.}},
\bauthor{\bsnm{Makida}, \binits{Y.}},
\bauthor{\bsnm{Mammei}, \binits{J.}},
\bauthor{\bsnm{Mammei}, \binits{R.}},
\bauthor{\bsnm{Martin}, \binits{J.W.}},
\bauthor{\bsnm{Matsumiya}, \binits{R.}},
\bauthor{\bsnm{Miller}, \binits{E.}},
\bauthor{\bsnm{Mishima}, \binits{K.}},
\bauthor{\bsnm{Momose}, \binits{T.}},
\bauthor{\bsnm{Okamura}, \binits{T.}},
\bauthor{\bsnm{Page}, \binits{S.}},
\bauthor{\bsnm{Picker}, \binits{R.}},
\bauthor{\bsnm{Pierre}, \binits{E.}},
\bauthor{\bsnm{Ramsay}, \binits{W.D.}},
\bauthor{\bsnm{Rebenitsch}, \binits{L.}},
\bauthor{\bsnm{Rehm}, \binits{F.}},
\bauthor{\bsnm{Schreyer}, \binits{W.}},
\bauthor{\bsnm{Shimizu}, \binits{H.M.}},
\bauthor{\bsnm{Sidhu}, \binits{S.}},
\bauthor{\bsnm{Sikora}, \binits{A.}},
\bauthor{\bsnm{Smith}, \binits{J.}},
\bauthor{\bsnm{Tanihata}, \binits{I.}},
\bauthor{\bsnm{Thorsteinson}, \binits{B.}},
\bauthor{\bsnm{Vanbergen}, \binits{S.}},
\bauthor{\bsnm{Oers}, \binits{W.T.H.}},
\bauthor{\bsnm{Watanabe}, \binits{Y.X.}}:
\batitle{{First ultracold neutrons produced at TRIUMF}}.
\bjtitle{Phys. Rev. C}
\bvolume{99},
\bfpage{025503}
(\byear{2019})
\doiurl{10.1103/PhysRevC.99.025503}
\end{barticle}
\endbibitem

\bibitem[\protect\citeauthoryear{Ahmed et~al.}{2019b}]{TUCAN_kicker}
\begin{barticle}
\bauthor{\bsnm{Ahmed}, \binits{S.}},
\bauthor{\bsnm{Altiere}, \binits{E.}},
\bauthor{\bsnm{Andalib}, \binits{T.}},
\bauthor{\bsnm{Barnes}, \binits{M.J.}},
\bauthor{\bsnm{Bell}, \binits{B.}},
\bauthor{\bsnm{Bidinosti}, \binits{C.P.}},
\bauthor{\bsnm{Bylinsky}, \binits{Y.}},
\bauthor{\bsnm{Chak}, \binits{J.}},
\bauthor{\bsnm{Das}, \binits{M.}},
\bauthor{\bsnm{Davis}, \binits{C.A.}},
\bauthor{\bsnm{Fischer}, \binits{F.}},
\bauthor{\bsnm{Franke}, \binits{B.}},
\bauthor{\bsnm{Gericke}, \binits{M.T.W.}},
\bauthor{\bsnm{Giampa}, \binits{P.}},
\bauthor{\bsnm{Hahn}, \binits{M.}},
\bauthor{\bsnm{Hansen-Romu}, \binits{S.}},
\bauthor{\bsnm{Hatanaka}, \binits{K.}},
\bauthor{\bsnm{Hayamizu}, \binits{T.}},
\bauthor{\bsnm{Jamieson}, \binits{B.}},
\bauthor{\bsnm{Jones}, \binits{D.}},
\bauthor{\bsnm{Katsika}, \binits{K.}},
\bauthor{\bsnm{Kawasaki}, \binits{S.}},
\bauthor{\bsnm{Kikawa}, \binits{T.}},
\bauthor{\bsnm{Klassen}, \binits{W.}},
\bauthor{\bsnm{Konaka}, \binits{A.}},
\bauthor{\bsnm{Korkmaz}, \binits{E.}},
\bauthor{\bsnm{Kuchler}, \binits{F.}},
\bauthor{\bsnm{Kurchaninov}, \binits{L.}},
\bauthor{\bsnm{Lang}, \binits{M.}},
\bauthor{\bsnm{Lee}, \binits{L.}},
\bauthor{\bsnm{Lindner}, \binits{T.}},
\bauthor{\bsnm{Madison}, \binits{K.W.}},
\bauthor{\bsnm{Mammei}, \binits{J.}},
\bauthor{\bsnm{Mammei}, \binits{R.}},
\bauthor{\bsnm{Martin}, \binits{J.W.}},
\bauthor{\bsnm{Matsumiya}, \binits{R.}},
\bauthor{\bsnm{Miller}, \binits{E.}},
\bauthor{\bsnm{Momose}, \binits{T.}},
\bauthor{\bsnm{Picker}, \binits{R.}},
\bauthor{\bsnm{Pierre}, \binits{E.}},
\bauthor{\bsnm{Ramsay}, \binits{W.D.}},
\bauthor{\bsnm{Rao}, \binits{Y.-N.}},
\bauthor{\bsnm{Rawnsley}, \binits{W.R.}},
\bauthor{\bsnm{Rebenitsch}, \binits{L.}},
\bauthor{\bsnm{Schreyer}, \binits{W.}},
\bauthor{\bsnm{Sidhu}, \binits{S.}},
\bauthor{\bsnm{Vanbergen}, \binits{S.}},
\bauthor{\bsnm{Oers}, \binits{W.T.H.}},
\bauthor{\bsnm{Watanabe}, \binits{Y.X.}},
\bauthor{\bsnm{Yosifov}, \binits{D.}}:
\batitle{{Fast-switching magnet serving a spallation-driven ultracold neutron source}}.
\bjtitle{Phys. Rev. Accel. Beams}
\bvolume{22},
\bfpage{102401}
(\byear{2019})
\doiurl{10.1103/PhysRevAccelBeams.22.102401}
\end{barticle}
\endbibitem

\bibitem[\protect\citeauthoryear{Ahmed et~al.}{2019c}]{TUCAN_beam}
\begin{barticle}
\bauthor{\bsnm{Ahmed}, \binits{S.}},
\bauthor{\bsnm{Andalib}, \binits{T.}},
\bauthor{\bsnm{Barnes}, \binits{M.J.}},
\bauthor{\bsnm{Bidinosti}, \binits{C.B.}},
\bauthor{\bsnm{Bylinsky}, \binits{Y.}},
\bauthor{\bsnm{Chak}, \binits{J.}},
\bauthor{\bsnm{Das}, \binits{M.}},
\bauthor{\bsnm{Davis}, \binits{C.A.}},
\bauthor{\bsnm{Franke}, \binits{B.}},
\bauthor{\bsnm{Gericke}, \binits{M.T.W.}},
\bauthor{\bsnm{Giampa}, \binits{P.}},
\bauthor{\bsnm{Hahn}, \binits{M.}},
\bauthor{\bsnm{Hansen-Romu}, \binits{S.}},
\bauthor{\bsnm{Hatanaka}, \binits{K.}},
\bauthor{\bsnm{Jamieson}, \binits{B.}},
\bauthor{\bsnm{Jones}, \binits{D.}},
\bauthor{\bsnm{Katsika}, \binits{K.}},
\bauthor{\bsnm{Kawasaki}, \binits{S.}},
\bauthor{\bsnm{Klassen}, \binits{W.}},
\bauthor{\bsnm{Konaka}, \binits{A.}},
\bauthor{\bsnm{Korkmaz}, \binits{E.}},
\bauthor{\bsnm{Kuchler}, \binits{F.}},
\bauthor{\bsnm{Kurchaninov}, \binits{L.}},
\bauthor{\bsnm{Lang}, \binits{M.}},
\bauthor{\bsnm{Lee}, \binits{L.}},
\bauthor{\bsnm{Lindner}, \binits{T.}},
\bauthor{\bsnm{Madison}, \binits{K.W.}},
\bauthor{\bsnm{Mammei}, \binits{J.}},
\bauthor{\bsnm{Mammei}, \binits{R.}},
\bauthor{\bsnm{Martin}, \binits{J.W.}},
\bauthor{\bsnm{Matsumiya}, \binits{R.}},
\bauthor{\bsnm{Picker}, \binits{R.}},
\bauthor{\bsnm{Pierre}, \binits{E.}},
\bauthor{\bsnm{Ramsay}, \binits{W.D.}},
\bauthor{\bsnm{Rao}, \binits{Y.-N.}},
\bauthor{\bsnm{Rawnsley}, \binits{W.R.}},
\bauthor{\bsnm{Rebenitsch}, \binits{L.}},
\bauthor{\bsnm{Remon}, \binits{C.A.}},
\bauthor{\bsnm{Schreyer}, \binits{W.}},
\bauthor{\bsnm{Sikora}, \binits{A.}},
\bauthor{\bsnm{Sidhu}, \binits{S.}},
\bauthor{\bsnm{Sonier}, \binits{J.}},
\bauthor{\bsnm{Thorsteinson}, \binits{B.}},
\bauthor{\bsnm{Vanbergen}, \binits{S.}},
\bauthor{\bsnm{{van Oers}}, \binits{W.T.H.}},
\bauthor{\bsnm{Watanabe}, \binits{Y.X.}},
\bauthor{\bsnm{Yosifov}, \binits{D.}}:
\batitle{{A beamline for fundamental neutron physics at TRIUMF}}.
\bjtitle{NIM-A}
\bvolume{927},
\bfpage{101}--\blpage{108}
(\byear{2019})
\doiurl{10.1016/j.nima.2019.01.074}
\end{barticle}
\endbibitem

\bibitem[\protect\citeauthoryear{Schreyer et~al.}{2020}]{TUCAN_moderator}
\begin{barticle}
\bauthor{\bsnm{Schreyer}, \binits{W.}},
\bauthor{\bsnm{Davis}, \binits{C.A.}},
\bauthor{\bsnm{Kawasaki}, \binits{S.}},
\bauthor{\bsnm{Kikawa}, \binits{T.}},
\bauthor{\bsnm{Marshall}, \binits{C.}},
\bauthor{\bsnm{Mishima}, \binits{K.}},
\bauthor{\bsnm{Okamura}, \binits{T.}},
\bauthor{\bsnm{Picker}, \binits{R.}}:
\batitle{{Optimizing neutron moderators for a spallation-driven ultracold-neutron source at TRIUMF}}.
\bjtitle{NIM-A}
\bvolume{959},
\bfpage{163525}
(\byear{2020})
\doiurl{10.1016/j.nima.2020.163525}
\end{barticle}
\endbibitem

\bibitem[\protect\citeauthoryear{Abel et~al.}{2020}]{abel2020}
\begin{barticle}
\bauthor{\bsnm{Abel}, \binits{C.}},
\bauthor{\bsnm{Afach}, \binits{S.}},
\bauthor{\bsnm{Ayres}, \binits{N.J.}},
\bauthor{\bsnm{Ban}, \binits{G.}},
\bauthor{\bsnm{Bison}, \binits{G.}},
\bauthor{\bsnm{Bodek}, \binits{K.}},
\bauthor{\bsnm{Bondar}, \binits{V.}},
\bauthor{\bsnm{Chanel}, \binits{E.}},
\bauthor{\bsnm{Chiu}, \binits{P.-J.}},
\bauthor{\bsnm{Crawford}, \binits{C.B.}},
\bauthor{\bsnm{Chowdhuri}, \binits{Z.}},
\bauthor{\bsnm{Daum}, \binits{M.}},
\bauthor{\bsnm{Emmenegger}, \binits{S.}},
\bauthor{\bsnm{Ferraris-Bouchez}, \binits{L.}},
\bauthor{\bsnm{Fertl}, \binits{M.}},
\bauthor{\bsnm{Franke}, \binits{B.}},
\bauthor{\bsnm{Griffith}, \binits{W.C.}},
\bauthor{\bsnm{Gruji\'{c}}, \binits{Z.D.}},
\bauthor{\bsnm{Hayen}, \binits{L.}},
\bauthor{\bsnm{H\'elaine}, \binits{V.}},
\bauthor{\bsnm{Hild}, \binits{N.}},
\bauthor{\bsnm{Kasprzak}, \binits{M.}},
\bauthor{\bsnm{Kermaidic}, \binits{Y.}},
\bauthor{\bsnm{Kirch}, \binits{K.}},
\bauthor{\bsnm{Knowles}, \binits{P.}},
\bauthor{\bsnm{Koch}, \binits{H.-C.}},
\bauthor{\bsnm{Komposch}, \binits{S.}},
\bauthor{\bsnm{Koss}, \binits{P.A.}},
\bauthor{\bsnm{Kozela}, \binits{A.}},
\bauthor{\bsnm{Krempel}, \binits{J.}},
\bauthor{\bsnm{Lauss}, \binits{B.}},
\bauthor{\bsnm{Lefort}, \binits{T.}},
\bauthor{\bsnm{Lemi\`ere}, \binits{Y.}},
\bauthor{\bsnm{Leredde}, \binits{A.}},
\bauthor{\bsnm{Mtchedlishvili}, \binits{A.}},
\bauthor{\bsnm{Mohanmurthy}, \binits{P.}},
\bauthor{\bsnm{Musgrave}, \binits{M.}},
\bauthor{\bsnm{Naviliat-Cuncic}, \binits{O.}},
\bauthor{\bsnm{Pais}, \binits{D.}},
\bauthor{\bsnm{Pazgalev}, \binits{A.}},
\bauthor{\bsnm{Piegsa}, \binits{F.M.}},
\bauthor{\bsnm{Pierre}, \binits{E.}},
\bauthor{\bsnm{Pignol}, \binits{G.}},
\bauthor{\bsnm{Prashanth}, \binits{P.N.}},
\bauthor{\bsnm{Qu\'em\'ener}, \binits{G.}},
\bauthor{\bsnm{Rawlik}, \binits{M.}},
\bauthor{\bsnm{Rebreyend}, \binits{D.}},
\bauthor{\bsnm{Ries}, \binits{D.}},
\bauthor{\bsnm{Roccia}, \binits{S.}},
\bauthor{\bsnm{Rozpedzik}, \binits{D.}},
\bauthor{\bsnm{Schmidt-Wellenburg}, \binits{P.}},
\bauthor{\bsnm{Schnabel}, \binits{A.}},
\bauthor{\bsnm{Severijns}, \binits{N.}},
\bauthor{\bsnm{Dinani}, \binits{R.T.}},
\bauthor{\bsnm{Thorne}, \binits{J.}},
\bauthor{\bsnm{Weis}, \binits{A.}},
\bauthor{\bsnm{Wursten}, \binits{E.}},
\bauthor{\bsnm{Wyszynski}, \binits{G.}},
\bauthor{\bsnm{Zejma}, \binits{J.}},
\bauthor{\bsnm{Zsigmond}, \binits{G.}}:
\batitle{{Optically pumped Cs magnetometers enabling a high-sensitivity search for the neutron electric dipole moment}}.
\bjtitle{Phys. Rev. A}
\bvolume{101},
\bfpage{053419}
(\byear{2020})
\doiurl{10.1103/PhysRevA.101.053419}
\end{barticle}
\endbibitem

\bibitem[\protect\citeauthoryear{Abel et~al.}{2022}]{psimapping_2022}
\begin{barticle}
\bauthor{\bsnm{Abel}, \binits{C.}},
\bauthor{\bsnm{Ayres}, \binits{N.J.}},
\bauthor{\bsnm{Ban}, \binits{G.}},
\bauthor{\bsnm{Bison}, \binits{G.}},
\bauthor{\bsnm{Bodek}, \binits{K.}},
\bauthor{\bsnm{Bondar}, \binits{V.}},
\bauthor{\bsnm{Chanel}, \binits{E.}},
\bauthor{\bsnm{Chiu}, \binits{P.-J.}},
\bauthor{\bsnm{Cl\'ement}, \binits{B.}},
\bauthor{\bsnm{Crawford}, \binits{C.B.}},
\bauthor{\bsnm{Daum}, \binits{M.}},
\bauthor{\bsnm{Emmenegger}, \binits{S.}},
\bauthor{\bsnm{Ferraris-Bouchez}, \binits{L.}},
\bauthor{\bsnm{Fertl}, \binits{M.}},
\bauthor{\bsnm{Flaux}, \binits{P.}},
\bauthor{\bsnm{Fratangelo}, \binits{A.}},
\bauthor{\bsnm{Griffith}, \binits{W.C.}},
\bauthor{\bsnm{Gruji\'{c}}, \binits{Z.D.}},
\bauthor{\bsnm{Harris}, \binits{P.G.}},
\bauthor{\bsnm{Hayen}, \binits{L.}},
\bauthor{\bsnm{Hild}, \binits{N.}},
\bauthor{\bsnm{Kasprzak}, \binits{M.}},
\bauthor{\bsnm{Kirch}, \binits{K.}},
\bauthor{\bsnm{Knowles}, \binits{P.}},
\bauthor{\bsnm{Koch}, \binits{H.-C.}},
\bauthor{\bsnm{Koss}, \binits{P.A.}},
\bauthor{\bsnm{Kozela}, \binits{A.}},
\bauthor{\bsnm{Krempel}, \binits{J.}},
\bauthor{\bsnm{Lauss}, \binits{B.}},
\bauthor{\bsnm{Lefort}, \binits{T.}},
\bauthor{\bsnm{Lemi\`ere}, \binits{Y.}},
\bauthor{\bsnm{Mohanmurthy}, \binits{P.}},
\bauthor{\bsnm{Naviliat-Cuncic}, \binits{O.}},
\bauthor{\bsnm{Pais}, \binits{D.}},
\bauthor{\bsnm{Piegsa}, \binits{F.M.}},
\bauthor{\bsnm{Pignol}, \binits{G.}},
\bauthor{\bsnm{Prashanth}, \binits{P.N.}},
\bauthor{\bsnm{Qu\'em\'ener}, \binits{G.}},
\bauthor{\bsnm{Rawlik}, \binits{M.}},
\bauthor{\bsnm{Ries}, \binits{D.}},
\bauthor{\bsnm{Rebreyend}, \binits{D.}},
\bauthor{\bsnm{Roccia}, \binits{S.}},
\bauthor{\bsnm{Rozpedzik}, \binits{D.}},
\bauthor{\bsnm{Schmidt-Wellenburg}, \binits{P.}},
\bauthor{\bsnm{Schnabel}, \binits{A.}},
\bauthor{\bsnm{Severijns}, \binits{N.}},
\bauthor{\bsnm{Thorne}, \binits{J.A.}},
\bauthor{\bsnm{Virot}, \binits{R.}},
\bauthor{\bsnm{Weis}, \binits{A.}},
\bauthor{\bsnm{Wursten}, \binits{E.}},
\bauthor{\bsnm{Wyszynski}, \binits{G.}},
\bauthor{\bsnm{Zejma}, \binits{J.}},
\bauthor{\bsnm{Zsigmond}, \binits{G.}}:
\batitle{{Mapping of the magnetic field to correct systematic effects in a neutron electric dipole moment experiment}}.
\bjtitle{Phys. Rev. A}
\bvolume{106},
\bfpage{032808}
(\byear{2022})
\doiurl{10.1103/PhysRevA.106.032808}
\end{barticle}
\endbibitem

\bibitem[\protect\citeauthoryear{Budker et~al.}{2002}]{Budker2002}
\begin{barticle}
\bauthor{\bsnm{Budker}, \binits{D.}},
\bauthor{\bsnm{Gawlik}, \binits{W.}},
\bauthor{\bsnm{Kimball}, \binits{D.F.}},
\bauthor{\bsnm{Rochester}, \binits{S.M.}},
\bauthor{\bsnm{Yashchuk}, \binits{V.V.}},
\bauthor{\bsnm{Weis}, \binits{A.}}:
\batitle{Resonant nonlinear magneto-optical effects in atoms}.
\bjtitle{Rev. Mod. Phys.}
\bvolume{74},
\bfpage{1153}--\blpage{1201}
(\byear{2002})
\doiurl{10.1103/RevModPhys.74.1153}
\end{barticle}
\endbibitem

\bibitem[\protect\citeauthoryear{Pustelny et~al.}{2008}]{Gawlik2008}
\begin{barticle}
\bauthor{\bsnm{Pustelny}, \binits{S.}},
\bauthor{\bsnm{Wojciechowski}, \binits{A.}},
\bauthor{\bsnm{Gring}, \binits{M.}},
\bauthor{\bsnm{Kotyrba}, \binits{M.}},
\bauthor{\bsnm{Zachorowski}, \binits{J.}},
\bauthor{\bsnm{Gawlik}, \binits{W.}}:
\batitle{{Magnetometry based on nonlinear magneto-optical rotation with amplitude-modulated light}}.
\bjtitle{Journal of Applied Physics}
\bvolume{103}(\bissue{6}),
\bfpage{063108}
(\byear{2008})
\doiurl{10.1063/1.2844494}
{\href{https://arxiv.org/abs/https://pubs.aip.org/aip/jap/article-pdf/doi/10.1063/1.2844494/13946272/063108\_1\_online.pdf}{{https://pubs.aip.org/aip/jap/article-pdf/doi/10.1063/1.2844494/13946272/063108\_1\_online.pdf}}}
\end{barticle}
\endbibitem

\bibitem[\protect\citeauthoryear{Das}{2018}]{Das_2018}
\begin{botherref}
\oauthor{\bsnm{Das}, \binits{M.}}:
Highly sensitive rb magnetometer for neutron electric dipole moment experiments.
Masters thesis,
University of Manitoba,
Winnipeg, Canada
(2018).
Available at \url{https://mspace.lib.umanitoba.ca/items/c95adc8a-6ba2-49db-9bbc-7e8e0bad5180}
\end{botherref}
\endbibitem

\bibitem[\protect\citeauthoryear{Klassen}{2020}]{klassen2020}
\begin{botherref}
\oauthor{\bsnm{Klassen}, \binits{W.}}:
{All-optical Cs Magnetometry system for a neutron electric dipole moment experiment}.
Masters thesis,
University of Manitoba,
Winnipeg, Canada
(March 2020).
Available at \url{https://mspace.lib.umanitoba.ca/items/5a0d92cc-be99-4df2-a102-e7f1ab0979b0}
\end{botherref}
\endbibitem

\bibitem[\protect\citeauthoryear{Higbie et~al.}{2006}]{higbie2006}
\begin{barticle}
\bauthor{\bsnm{Higbie}, \binits{J.M.}},
\bauthor{\bsnm{Corsini}, \binits{E.}},
\bauthor{\bsnm{Budker}, \binits{D.}}:
\batitle{{Robust, high-speed, all-optical atomic magnetometer}}.
\bjtitle{Rev. Sci. Instrum.}
\bvolume{77}(\bissue{11}),
\bfpage{113106}
(\byear{2006})
\doiurl{10.1063/1.2370597}
{\href{https://arxiv.org/abs/https://pubs.aip.org/aip/rsi/article-pdf/doi/10.1063/1.2370597/8816016/113106\_1\_online.pdf}{{https://pubs.aip.org/aip/rsi/article-pdf/doi/10.1063/1.2370597/8816016/113106\_1\_online.pdf}}}
\end{barticle}
\endbibitem

\bibitem[\protect\citeauthoryear{Bell and Bloom}{1961}]{bellbloom}
\begin{barticle}
\bauthor{\bsnm{Bell}, \binits{W.E.}},
\bauthor{\bsnm{Bloom}, \binits{A.L.}}:
\batitle{{Optically Driven Spin Precession}}.
\bjtitle{Phys. Rev. Lett.}
\bvolume{6},
\bfpage{280}--\blpage{281}
(\byear{1961})
\doiurl{10.1103/PhysRevLett.6.280}
\end{barticle}
\endbibitem

\bibitem[\protect\citeauthoryear{Gruji{\'{c}} et~al.}{2015}]{Grujic_2015}
\begin{barticle}
\bauthor{\bsnm{Gruji{\'{c}}}, \binits{Z.D.}},
\bauthor{\bsnm{Koss}, \binits{P.A.}},
\bauthor{\bsnm{Bison}, \binits{G.}},
\bauthor{\bsnm{Weis}, \binits{A.}}:
\batitle{{A sensitive and accurate atomic magnetometer based on free spin precession}}.
\bjtitle{EPJD}
\bvolume{69}(\bissue{5}),
\bfpage{135}
(\byear{2015})
\doiurl{10.1140/epjd/e2015-50875-3}
\end{barticle}
\endbibitem

\bibitem[\protect\citeauthoryear{Budker et~al.}{2013}]{budker2013}
\begin{bchapter}
\bauthor{\bsnm{Budker}, \binits{D.}},
\bauthor{\bsnm{Kimball}, \binits{J.D.F.}},
\bauthor{\bsnm{Pustelny}, \binits{S.}},
\bauthor{\bsnm{Yashchuk}, \binits{V.V.}}:
\bctitle{Optical magnetometry with modulated light}.
In: \beditor{\bsnm{Budker}, \binits{D.}},
\beditor{\bsnm{Kimball}, \binits{J.D.F.}} (eds.)
\bbtitle{Optical Magnetometry},
pp. \bfpage{104}--\blpage{122}.
\bpublisher{Cambridge University Press},
\blocation{Cambridge, UK}
(\byear{2013}).
\bcomment{Chap. 6}
\end{bchapter}
\endbibitem

\bibitem[\protect\citeauthoryear{Rosner et~al.}{2022}]{rosner2022}
\begin{barticle}
\bauthor{\bsnm{Rosner}, \binits{M.}},
\bauthor{\bsnm{Beck}, \binits{D.}},
\bauthor{\bsnm{Fierlinger}, \binits{P.}},
\bauthor{\bsnm{Filter}, \binits{H.}},
\bauthor{\bsnm{Klau}, \binits{C.}},
\bauthor{\bsnm{Kuchler}, \binits{F.}},
\bauthor{\bsnm{Rößner}, \binits{P.}},
\bauthor{\bsnm{Sturm}, \binits{M.}},
\bauthor{\bsnm{Wurm}, \binits{D.}},
\bauthor{\bsnm{Sun}, \binits{Z.}}:
\batitle{{A highly drift-stable atomic magnetometer for fundamental physics experiments}}.
\bjtitle{Appl. Phys. Lett.}
\bvolume{120}(\bissue{16}),
\bfpage{161102}
(\byear{2022})
\doiurl{10.1063/5.0083854}
{\href{https://arxiv.org/abs/https://doi.org/10.1063/5.0083854}{{https://doi.org/10.1063/5.0083854}}}
\end{barticle}
\endbibitem

\bibitem[\protect\citeauthoryear{Rosner}{2022}]{Rosner2022thesis}
\begin{botherref}
\oauthor{\bsnm{Rosner}, \binits{M.}}:
Drift-stable magnetometry for fundamental physics experiments.
Phd. thesis,
Munich, Tech. U.,
Munich, Germany
(2022).
Available at \url{https://mediatum.ub.tum.de/1654045?show_id=1658114}
\end{botherref}
\endbibitem

\bibitem[\protect\citeauthoryear{Sturm}{2020}]{Sturm:2020gip}
\begin{botherref}
\oauthor{\bsnm{Sturm}, \binits{M.}}:
{A highly drift stable and fully optical Cs atomic magnetometer for a new generation nEDM experiment}.
Phd. thesis,
Munich, Tech. U.,
Munich, Germany
(2020).
Available at \url{https://mediatum.ub.tum.de/?id=1536319}
\end{botherref}
\endbibitem

\bibitem[\protect\citeauthoryear{}{2016}]{SouthwestSciences_2016}
\begin{botherref}
(2016).
\url{https://www.swsciences.com/contact/index.html}
\end{botherref}
\endbibitem

\bibitem[\protect\citeauthoryear{{Precision Glassblowing of Colorado, Inc.}}{2024}]{PGB2024}
\begin{botherref}
\oauthor{\bsnm{{Precision Glassblowing of Colorado, Inc.}}}:
About precision glassblowing
(2024).
\url{https://precisionglassblowing.com/about-precision-glassblowing/}
\end{botherref}
\endbibitem

\bibitem[\protect\citeauthoryear{}{}]{Summersoptical}
\begin{botherref}
\url{https://www.optical-cement.com/default.htm}
\end{botherref}
\endbibitem

\bibitem[\protect\citeauthoryear{{D-TACQ Solutions Ltd}}{}]{D-TACQ}
\begin{botherref}
\oauthor{\bsnm{{D-TACQ Solutions Ltd}}}:
Intelligent Data Acquisition Boards and Systems.
\url{https://www.d-tacq.com/}
\end{botherref}
\endbibitem

\bibitem[\protect\citeauthoryear{Martin et~al.}{2015}]{MARTIN2015}
\begin{barticle}
\bauthor{\bsnm{Martin}, \binits{J.W.}},
\bauthor{\bsnm{Mammei}, \binits{R.R.}},
\bauthor{\bsnm{Klassen}, \binits{W.}},
\bauthor{\bsnm{Cerasani}, \binits{C.}},
\bauthor{\bsnm{Andalib}, \binits{T.}},
\bauthor{\bsnm{Bidinosti}, \binits{C.P.}},
\bauthor{\bsnm{Lang}, \binits{M.}},
\bauthor{\bsnm{Ostapchuk}, \binits{D.}}:
\batitle{{Large magnetic shielding factor measured by nonlinear magneto-optical rotation}}.
\bjtitle{NIM-A}
\bvolume{778},
\bfpage{61}--\blpage{66}
(\byear{2015})
\doiurl{10.1016/j.nima.2015.01.003}
\end{barticle}
\endbibitem

\bibitem[\protect\citeauthoryear{Ahmed et~al.}{2024}]{Ahmed24}
\begin{bchapter}
\bauthor{\bsnm{Ahmed}, \binits{S.}},
\bauthor{\bsnm{Jamieson}, \binits{B.}},
\bauthor{\bsnm{Martin}, \binits{J.W.}},
\bauthor{\bsnm{Ostapchuk}, \binits{D.C.M.}},
\bauthor{\bsnm{McCrea}, \binits{M.}},
\bauthor{\bsnm{Klassen}, \binits{W.}}:
\bctitle{An {Ultra-Stable} custom current supply for use in a neutron electric dipole moment experiment}.
In: \bbtitle{2024 IEEE 22nd Mediterranean Electrotechnical Conference (MELECON) (MELECON 2024)},
\bconflocation{Porto, Portugal},
p. \bfpage{5}
(\byear{2024})
\end{bchapter}
\endbibitem

\bibitem[\protect\citeauthoryear{Gemmel et~al.}{2010}]{Gemmel2010}
\begin{barticle}
\bauthor{\bsnm{Gemmel}, \binits{C.}},
\bauthor{\bsnm{Heil}, \binits{W.}},
\bauthor{\bsnm{Karpuk}, \binits{S.}},
\bauthor{\bsnm{Lenz}, \binits{K.}},
\bauthor{\bsnm{Ludwig}, \binits{C.}},
\bauthor{\bsnm{Sobolev}, \binits{Y.}},
\bauthor{\bsnm{Tullney}, \binits{K.}},
\bauthor{\bsnm{Burghoff}, \binits{M.}},
\bauthor{\bsnm{Kilian}, \binits{W.}},
\bauthor{\bsnm{Knappe-Gr{\"u}neberg}, \binits{S.}},
\bauthor{\bsnm{M{\"u}ller}, \binits{W.}},
\bauthor{\bsnm{Schnabel}, \binits{A.}},
\bauthor{\bsnm{Seifert}, \binits{F.}},
\bauthor{\bsnm{Trahms}, \binits{L.}},
\bauthor{\bsnm{Bae{\ss}ler}, \binits{S.}}:
\batitle{Ultra-sensitive magnetometry based on free precession of nuclear spins}.
\bjtitle{EPJD}
\bvolume{57}(\bissue{3}),
\bfpage{303}--\blpage{320}
(\byear{2010})
\doiurl{10.1140/epjd/e2010-00044-5}
\end{barticle}
\endbibitem

\end{thebibliography}

\end{document}